\documentclass[twoside,twocolumn,9pt]{article}
\usepackage{extsizes}
\usepackage[super,sort&compress,comma]{natbib} 
\usepackage[version=3]{mhchem}
\usepackage[left=1.5cm, right=1.5cm, top=1.785cm, bottom=2.0cm]{geometry}
\usepackage{balance}
\usepackage{mathptmx}
\usepackage{sectsty}
\usepackage{graphicx} 
\usepackage{lastpage}
\usepackage{amssymb}
\usepackage[format=plain,justification=justified,singlelinecheck=false,font={stretch=1.125,small,sf},labelfont=bf,labelsep=space]{caption}
\usepackage{float}
\usepackage{fancyhdr}
\usepackage{fnpos}
\usepackage[english]{babel}
\addto{\captionsenglish}{%
	
}
\usepackage{array}
\usepackage{droidsans}
\usepackage{charter}
\usepackage[T1]{fontenc}
\usepackage[usenames,dvipsnames]{xcolor}
\usepackage{setspace}
\usepackage[compact]{titlesec}
\usepackage{hyperref}
\usepackage{bm} 
\usepackage{caption}
\usepackage{subcaption}
\usepackage{epstopdf}

\definecolor{cream}{RGB}{222,217,201}

\begin{document}
	
	\pagestyle{fancy}
	\thispagestyle{plain}
	\fancypagestyle{plain}{
		\renewcommand{\headrulewidth}{0pt}
	}
	
	\makeFNbottom
	\makeatletter
	\renewcommand\LARGE{\@setfontsize\LARGE{15pt}{17}}
	\renewcommand\Large{\@setfontsize\Large{12pt}{14}}
	\renewcommand\large{\@setfontsize\large{10pt}{12}}
	\renewcommand\footnotesize{\@setfontsize\footnotesize{7pt}{10}}
	\makeatother
	
	\renewcommand{\thefootnote}{\fnsymbol{footnote}}
	\renewcommand\footnoterule{\vspace*{1pt}%
		\color{cream}\hrule width 3.5in height 0.4pt \color{black}\vspace*{5pt}} 
	\setcounter{secnumdepth}{5}
	
	\makeatletter 
	\renewcommand\@biblabel[1]{#1}            
	\renewcommand\@makefntext[1]%
	{\noindent\makebox[0pt][r]{\@thefnmark\,}#1}
	\makeatother 
	\renewcommand{\figurename}{\small{Fig.}~}
	\sectionfont{\sffamily\Large}
	\subsectionfont{\normalsize}
	\subsubsectionfont{\bf}
	\setstretch{1.125} 
	\setlength{\skip\footins}{0.8cm}
	\setlength{\footnotesep}{0.25cm}
	\setlength{\jot}{10pt}
	\titlespacing*{\section}{0pt}{4pt}{4pt}
	\titlespacing*{\subsection}{0pt}{15pt}{1pt}
	
	\fancyfoot{}
	\fancyfoot[LO,RE]{\vspace{-7.1pt}\includegraphics[height=9pt]{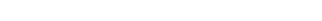}}
	\fancyfoot[CO]{\vspace{-7.1pt}\hspace{13.2cm}\includegraphics{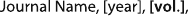}}
	\fancyfoot[CE]{\vspace{-7.2pt}\hspace{-14.2cm}\includegraphics{head_foot/RF}}
	\fancyfoot[RO]{\footnotesize{\sffamily{1--\pageref{LastPage} ~\textbar  \hspace{2pt}\thepage}}}
	\fancyfoot[LE]{\footnotesize{\sffamily{\thepage~\textbar\hspace{3.45cm} 1--\pageref{LastPage}}}}
	\fancyhead{}
	\renewcommand{\headrulewidth}{0pt} 
	\renewcommand{\footrulewidth}{0pt}
	\setlength{\arrayrulewidth}{1pt}
	\setlength{\columnsep}{6.5mm}
	\setlength\bibsep{1pt}
	
	\makeatletter 
	\newlength{\figrulesep} 
	\setlength{\figrulesep}{0.5\textfloatsep} 
	
	\newcommand{\topfigrule}{\vspace*{-1pt}%
		\noindent{\color{cream}\rule[-\figrulesep]{\columnwidth}{1.5pt}} }
	
	\newcommand{\botfigrule}{\vspace*{-2pt}%
		\noindent{\color{cream}\rule[\figrulesep]{\columnwidth}{1.5pt}} }
	
	\newcommand{\dblfigrule}{\vspace*{-1pt}%
		\noindent{\color{cream}\rule[-\figrulesep]{\textwidth}{1.5pt}} }
	
	\makeatother
	
	\twocolumn[
	\begin{@twocolumnfalse}
		\begin{tabular}{m{4.5cm} p{13.5cm} }
			
			 	\includegraphics{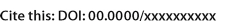} &\noindent\LARGE{\textbf{Oxygen Deficiency Drives Drastic Pattern Transition in Algal Bioconvection}} \\
			\vspace{0.3cm} & \vspace{0.3cm} \\
			
			& \noindent\large{Sangram Gore$^{\text{a}\S}$, Iraj Gholami$^{\text{a}\S}$, Samar O. Ahmed$^\text{a}$, Tomiris Doskhozhina$^\text{a}$, Sai V.R. Ambadipudi$^\text{a}$, Albert J. Bae$^\text{b}$ and Azam Gholami$^{\text{a}*}$} \\
			
			\includegraphics{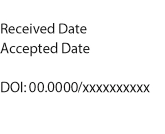} & \noindent\normalsize{
				\section*{Abstract}
				Suspensions of motile microorganisms can spontaneously form large-scale fluid motion, known as bioconvection, characterized by dense downwelling plumes separated by broad upwelling regions. In this study, we investigate bioconvection in shallow suspensions of \textit{Chlamydomonas reinhardtii} confined within spiral-shaped boundaries, combining detailed experiments with three-dimensional simulations. Under open liquid–air interfaces, cells accumulate near the surface via negative gravitaxis, generating spiral-shaped density patterns that subsequently fragment into lattice-like clusters, leading to plume formation. Space–time analyses demonstrate coherent rotational dynamics, with predominantly inward-directed motion near the spiral core and bidirectional motion further out. Introducing confinement by sealing the upper boundary with an air-impermeable wall triggers dramatic pattern transitions due to oxygen depletion: initially stable arrangements reorganize into new structures with significantly reduced wavelengths. Complementary numerical simulations, based on incompressible Navier–Stokes equations incorporating negative buoyancy and active swimmer stress, successfully replicate initial pattern formation, subsequent instability, fragmentation into plumes, and emergence of strong vortical flows—nearly an order of magnitude faster than individual cell swimming. However, these models do not capture oxygen depletion-driven transitions observed experimentally. Our results highlight that geometric confinement, oxygen availability, and metabolic transitions critically regulate bioconvection dynamics, offering novel strategies for controlling microbial self-organization and fluid transport.} \\
			
		\end{tabular}
		
	\end{@twocolumnfalse} \vspace{0.6cm}
	
	]
	
	\renewcommand*\rmdefault{bch}\normalfont\upshape
	\rmfamily
	\section*{}
	\vspace{-1cm}


	\footnotetext{$^\text{a}$New York University Abu Dhabi, Abu Dhabi, United Arab Emirates. $^\text{b}$ Department of Physics, Lewis \& Clark College, Portland, Oregon, USA.  \\E-mail: azam.gholami@nyu.edu }
	\footnotetext[4]{These authors contributed equally to this work.}
	\footnotetext{\dag~Electronic Supplementary Information (ESI) available: See DOI: 00.0000/00000000.}
	

	

\section{Introduction}
Microswimmers such as \textit{Escherichia coli}, \textit{Bacillus subtilis}, and the freshwater alga \textit{Chlamydomonas reinhardtii} generate characteristic fluid flows through the coordinated beating of their flagella and body movements~\cite{goldstein2015green,brumley2014flagellar,wan2016coordinated,drescher2011fluid,polin2009chlamydomonas,ahmad2021light,mojiri2021rapid}. Depending on their flagellar arrangement, these microorganisms are typically categorized as either "pushers" or "pullers"~\cite{pedley1987orientation,Pedley1992,drescher2011fluid,klindt2015flagellar,lauga2016bacterial}. In pusher-type swimmers, such as \textit{E. coli}, the thrust generated by flagellar motion propels fluid outward along the swimming axis and draws it inward from the sides, creating extensile flow fields. Conversely, puller-type swimmers, such as \textit{Chlamydomonas reinhardtii}, use anterior flagella to pull themselves forward, drawing fluid inward along the swimming axis and pushing it outward laterally, producing contractile flow fields. These microscale hydrodynamic disturbances significantly influence the motion of nearby passive particles and neighboring microorganisms, leading to complex collective behaviors, especially in dense suspensions. In particular, when large populations of motile microorganisms interact through the flows they generate, they can spontaneously organize into coherent large-scale structures, producing macroscopic flows through a process known as bioconvection. Unlike thermally driven convection, bioconvection arises solely from the active motility and orientation biases of the cells~\cite{Pedley1992,kessler1985hydrodynamic,bees1998linear,bees2020advances,hill2005bioconvection,pedley1990new,kessler1985co}, and is characterized by the emergence of dense, cell-rich downwelling plumes interspersed with broad regions of upwelling fluid.

Bioconvection is commonly observed in shallow laboratory suspensions, where cells swim upward on average due to directional responses known as "taxes," including gravitaxis (orientation against gravity), gyrotaxis (a balance between gravitational and viscous torques), phototaxis (response to light), and chemotaxis (response to chemical gradients)~\cite{bees1997wavelengths,williams2011tale,hill2005bioconvection,williams2011tale,Pedley1992,kessler1985hydrodynamic,javadi2020photo,dervaux2017light,ramamonjy2023pattern}. Gravitaxis, in particular, is typically a passive orientation mechanism resulting from physical asymmetry within the cells, such as bottom-heavy mass distribution~\cite{kessler1985hydrodynamic,kage2020shape,kessler1986individual}. Cells swimming upward due to negative gravitaxis accumulate near the upper suspension boundary, creating an unstable density gradient—since they are approximately 5$\%$ denser than water—that leads to the emergence of bioconvection patterns~\cite{bees1998linear,Pedley1992,hill2005bioconvection}. The fundamental mechanisms underlying bioconvection have been extensively modeled through continuum approaches~\cite{pedley1990new,childress1975pattern,pedley1987orientation}, incorporating momentum and mass conservation equations coupled with equations governing the orientation and distribution of swimming cells. Early work by Kessler~\cite{kessler1985hydrodynamic,kessler1986individual} and later developments by Pedley and Kessler~\cite{Pedley1992} established the theoretical framework for understanding how gyrotaxis and bottom-heaviness drive these large-scale convective phenomena.

Recent studies have demonstrated that cell geometry, fluid depth, and boundary conditions critically influence the formation and stability of bioconvection patterns~\cite{bees1997wavelengths,czirok2000bioconvective,czirok2000collective,bearon2011spatial,bearon2015trapping,karimi2013bioconvection}. While extensive research has explored bioconvection in simple geometries, the effects of more intricate internal structures, such as spiral-shaped boundaries, remain largely unexplored. Spiral geometries are particularly intriguing because they introduce geometrical singularities—regions of sharp curvature and varying local confinement—that can profoundly modulate cell accumulation, flow patterns, and instabilities. Additionally, the presence or absence of an open fluid–air interface significantly impacts bioconvection dynamics, primarily by regulating oxygen exchange between the suspension and the environment. Recent work by Fragkopoulos \textit{et al.}\cite{fragkopoulos2025metabolic} has shown that metabolic activity—closely linked to both light intensity and oxygen availability—directly modulates microbial motility and fundamentally governs the emergence and stability of coherent flow structures. These findings underscore the pivotal role of metabolic constraints, associated with light and oxygen dynamics, in shaping the spatiotemporal evolution of bioconvection patterns\cite{bees1998linear,williams2011tale,fragkopoulos2025metabolic,fragkopoulos2021self}.

In this study, we investigate the formation and evolution of bioconvection patterns in suspensions of \textit{Chlamydomonas reinhardtii} confined within spiral-shaped boundaries, under controlled red-light illumination to minimize phototactic effects. By systematically varying the boundary conditions—from open liquid–air interfaces to fully confined, air-impermeable geometries—we investigate how geometric confinement and oxygen depletion shape the spatio-temporal organization and characteristic wavelengths of bioconvection patterns. Through a combination of detailed experimental approaches—including top-view imaging, space–time analyses, and Fourier transforms—we show that confinement and oxygen depletion induce pronounced transitions in pattern morphology, resulting in the spontaneous reorganization of bioconvection patterns under oxygen-limited conditions. Complementary three-dimensional numerical simulations, based on the incompressible Navier–Stokes equations coupled with active-swimmer models, successfully reproduce the primary stages of pattern formation and hydrodynamic instability. However, these simulations do not capture the oxygen-dependent transitions observed experimentally. Together, our findings demonstrate that variations in confinement geometry and metabolic conditions can significantly alter collective microbial behavior, underscoring a strong interplay between microscale motility, environmental constraints, and emergent fluid flows.
\section{Materials and Methods} 
\subsection{Cell culture}
\textit{Chlamydomonas reinhardtii} (\textit{CR}) cultures were prepared following established protocols optimized to maintain high cell motility and physiological activity~\cite{harris2009chlamydomonas,catalan2025preparation}. Wild-type \textit{CR} cells (strain CC-409), obtained from the Algae Culture Collection at the University of Minnesota (USA), were initially cultured on agar plates containing Tris-Acetate-Phosphate (TAP) medium. The cultures were maintained under controlled illumination in a Memmert IPP260ecoplus incubator. To prepare liquid cultures, a small volume of cells from agar plates was inoculated into 500\,mL Erlenmeyer flasks containing TAP medium and incubated at 24\,$^\circ$C on an orbital shaker (INFORS HT Multitron) at 80\,rpm under continuous illumination with warm white LEDs. Vegetative cells were harvested during the logarithmic growth phase, typically on the third or fourth day after inoculation. Cell densities were measured using an automated cell counter (PerkinElmer Cellometer) and used at the final concentration of approximately 4$\times$10$^{6}$\,cells/mL.
\subsection{Experimental setup}
To investigate the influence of curved boundaries on bioconvection patterns in \textit{CR} suspensions, we fabricated molds from acrylic (PMMA)—a material impermeable to air—with different internal geometries, including spiral-shaped and circular designs. Each design was examined under either open (aerobic) or sealed (anaerobic) top boundary conditions to systematically probe the effects of oxygen availability. The experimental setup was carefully leveled and filled with a thin layer of cell suspension (typically 5\,mm deep) at a concentration of approximately 4$\times$10$^{6}$\,cells/mL. All experiments were conducted inside a dark enclosure to eliminate phototactic responses triggered by ambient light. To further suppress phototactic effects during imaging and pattern development, the suspension was illuminated exclusively from above using a red LED light source centered at 625 nm (LUMIMAX, LSR24-R), with an optical diffuser positioned between the light source and the sample to ensure uniform illumination. Previous studies have demonstrated that the phototactic sensitivity of \textit{CR} cells declines sharply at longer red wavelengths~\cite{berthold2008channelrhodopsin,witman1993chlamydomonas,ruffer1990flagellar}. Therefore, illumination at red light effectively eliminates phototactic steering, ensuring that the observed bioconvection patterns primarily result from hydrodynamic instabilities and gravitactic behavior.
\subsection{Image capture and analysis}
For each experiment, more than 1,000 top-view images were captured at 1-second intervals with an 8-bit depth. In experiments requiring longer observation, images were recorded continuously for up to two hours at a frame rate of 1 fps. The red LED illumination intensity, measured at the location of the PMMA mold, was maintained at approximately $5\times10^{17}$ photons\,m$^{-2}$\,s$^{-1}$. In the raw images, regions of high cell density appeared dark, whereas regions of low cell density appeared bright. To quantify the dominant wavelength of the emerging bioconvection patterns, Fourier analysis was performed on space–time plots using custom MATLAB scripts. Prior to analysis, image preprocessing was performed using FIJI. Initially, the images were inverted in intensity, and the mean image was subtracted from the stack to enhance the signal-to-noise ratio. In the resulting post-processed images, regions of high cell density appeared bright (visualized in pseudo-green), whereas regions of low density appeared dark.
\subsection{Model}
Bioconvection, the spontaneous formation of large-scale patterns in suspensions of swimming microorganisms, arises from the interplay between individual cell motility, fluid mechanics, and external fields such as gravity~\cite{childress1975pattern, pedley1992hydrodynamic,  hill2005bioconvection,bees2020advances,pedley1988growth,pedley2010collective,pedley1987orientation}. Most theoretical models of bioconvection aim to link macroscopic flow fields to microscopic swimming behaviors, often under the assumption that direct cell-cell interactions are negligible due to the dilute nature of the suspension. In these systems, the Reynolds number associated with individual swimmers is extremely low (around $10^{-3}$), reflecting viscous-dominated motion at the microscale. However, the collective flows generated by bioconvection can reach much higher Reynolds numbers, often on the order of $10$, highlighting a dramatic amplification of fluid motion through coordinated swimming~\cite{bearon2011spatial, pedley1987orientation,bees2020advances,drescher2011fluid}.

We model the collective dynamics of a dilute suspension of \textit{CR} cells using a continuum framework based on earlier studies~\cite{childress1975pattern, pedley1987orientation, pedley1990new}. The model couples the incompressible Navier--Stokes equations with a conservation equation for the cell concentration and an orientation dynamics equation accounting for gyrotaxis:
\begin{align}
	&\nabla \cdot \bm{U} = 0, \label{Eq:incompressibility} \\
	&\rho_0 \left( \frac{\partial \bm{U}}{\partial t} + (\bm{U} \cdot \nabla) \bm{U} \right) = -\nabla P - c g \Delta \rho \, \hat{z} + \nabla \cdot (2\mu \bm{E} + \bm{\Sigma}), \label{Eq:NS} \\
	&\frac{\partial c}{\partial t} = -\nabla \cdot \left[ c \left( \bm{U} + W_c \bm{p} - W_s \hat{z} \right) - D \nabla c \right], \label{Eq:Density} \\
	&\frac{\partial \bm{p}}{\partial t} = \frac{1}{2B} \left[ \hat{z} - (\hat{z} \cdot \bm{p})\bm{p} \right] + \frac{1}{2} \bm{\omega} \times \bm{p}, \label{gyrotaxis}
\end{align}
where $\bm{U}$ is the fluid velocity, $\bm{\omega}=\nabla\times \bm{U}$ is the vorticity field, $\mu$ is the fluid viscosity, $\rho_0$ is the fluid density, $P$ is the pressure, $c$ is the local cell concentration, and $\bm{p}$ denotes the local swimming direction of the cells. The stress tensor $\bm{E}=(\nabla \bm U+\nabla \bm U^{T})/2$ represents the rate of strain, and $\bm{\Sigma}$ is the active stress generated by the swimming cells, modeled as $\bm{\Sigma} = cS(\langle \bm{p}\bm{p} \rangle - \bm{I}/3)$~\cite{pedley2010instability,pedley1987orientation,pedley1990new,drescher2011fluid,aditi2002hydrodynamic}. Here $\bm I$ is the unity tensor, $S$ is the stresslet strength and is positive for \textit{CR} cells which are 'pullers' and negative for sperm cells which are pushers. The buoyancy term accounts for the density difference $\Delta \rho=\rho-\rho_0$ between cells and the surrounding fluid, and the Boussinesq approximation is employed~\cite{childress1975pattern,pedley1990new,bees2020advances}. Given that the characteristic timescale of bioconvection is considerably shorter than that of cellular growth, the conservation of the cell population is described by Eq.~\ref{Eq:Density}. This equation incorporates advection by both the fluid velocity and the active swimming motion at speed $W_c$, sedimentation under gravity at velocity $W_s$, and random motility modeled as an effective diffusion process with diffusivity $D$~\cite{pedley1990new,pedley1987orientation, bearon2011spatial}. Orientation dynamics (Eq.~\ref{gyrotaxis}) describes the balance between gravitational torque, which favors upward orientation, and viscous torque arising from fluid shear, resulting in gyrotactic reorientation over a characteristic timescale given by
$B = \alpha_\perp/2 h m g,$
where $\alpha_\perp = 8\pi\mu a^3$ is the rotational drag coefficient for a spherical cell of radius $a$, $h$ is the center-of-gravity offset within the cell, $m$ is the cell mass, and $g$ is the gravitational acceleration~\cite{pedley1987orientation, kessler1985hydrodynamic}.
This framework assumes the suspension is sufficiently dilute such that cell-cell hydrodynamic or steric interactions are negligible. Each cell is modeled as a sphere of volume $\beta$, swimming along its axis $\bm{p}$. Table~\ref{tab:parameters} summarizes the parameters used in the numerical simulations, selected based on experimental measurements of \textit{CR} cells and previous bioconvection studies~\cite{hill2005bioconvection, drescher2011fluid}. Numerical solutions of the coupled system were obtained using COMSOL Multiphysics (version 6.3), enabling us to reproduce some of the key features of the experimentally observed bioconvection patterns.
\begin{table}[htbp]
	\centering
	\caption{Parameters used in simulations of bioconvection patterns.}
	\label{tab:parameters}
	\begin{tabular}{l r}
		\hline\hline
		Mean concentration, $\bar c$                    & $4\times10^{6}\,\mathrm{cells/mL}$      \\
		Fluid depth, $L$                                & $5\,\mathrm{mm}$                              \\
		Fluid density, $\rho_0$                         & $997\,\mathrm{kg\,m^{-3}}$                    \\
		Cell density, $\rho$                            & $1050\,\mathrm{kg\,m^{-3}}$                   \\
		Cell radius, $a$                                & $5\times10^{-6}\,\mathrm{m}$                  \\
		Center of gravity offset, $h$                   & $10^{-5}\,\mathrm{cm}$                        \\
		Volume per cell, $\beta$                        & $5\times10^{-16}\,\mathrm{m^{3}}$             \\
		Swimming speed, $W_c$                           & $10^{-4}\,\mathrm{m\,s^{-1}}$                 \\
		Sedimentation speed, $W_s$                      & $6\times10^{-6}\,\mathrm{m\,s^{-1}}$          \\
		Cell diffusivity, $D$                           & $5\times10^{-8}\,\mathrm{m^{2}\,s^{-1}}$      \\
		Kinematic viscosity, $\nu$                      & $8.9\times10^{-7}\,\mathrm{m^{2}\,s^{-1}}$    \\
		Gyrotactic parameter, $B$                       & $1\,\mathrm{s}$                               \\
		Scaled swimming speed, $V_c = W_cL/D$           & $10$                                          \\
		Schmidt number, $S_c = \nu/D$                    & $17.85$                                       \\
		Gyrotaxis number, $G = BD/L^{2}$                & $2\times10^{-3}$                              \\
		Pseudo‐Rayleigh number, 
		$R_a = \dfrac{\bar c\,\beta\,\Delta\rho\,g\,L^{3}}{\rho\,\nu\,D}$ 
		& $2920$                                        \\
		\hline\hline
	\end{tabular}
\end{table}
\section{Results}
\subsection{Spiral-shaped boundaries induce large-scale rotational patterns}
\begin{figure*}[t]
	\centering
	\begin{subfigure}[c]{0.488\textwidth}
		\centering
		\includegraphics[width=\textwidth]{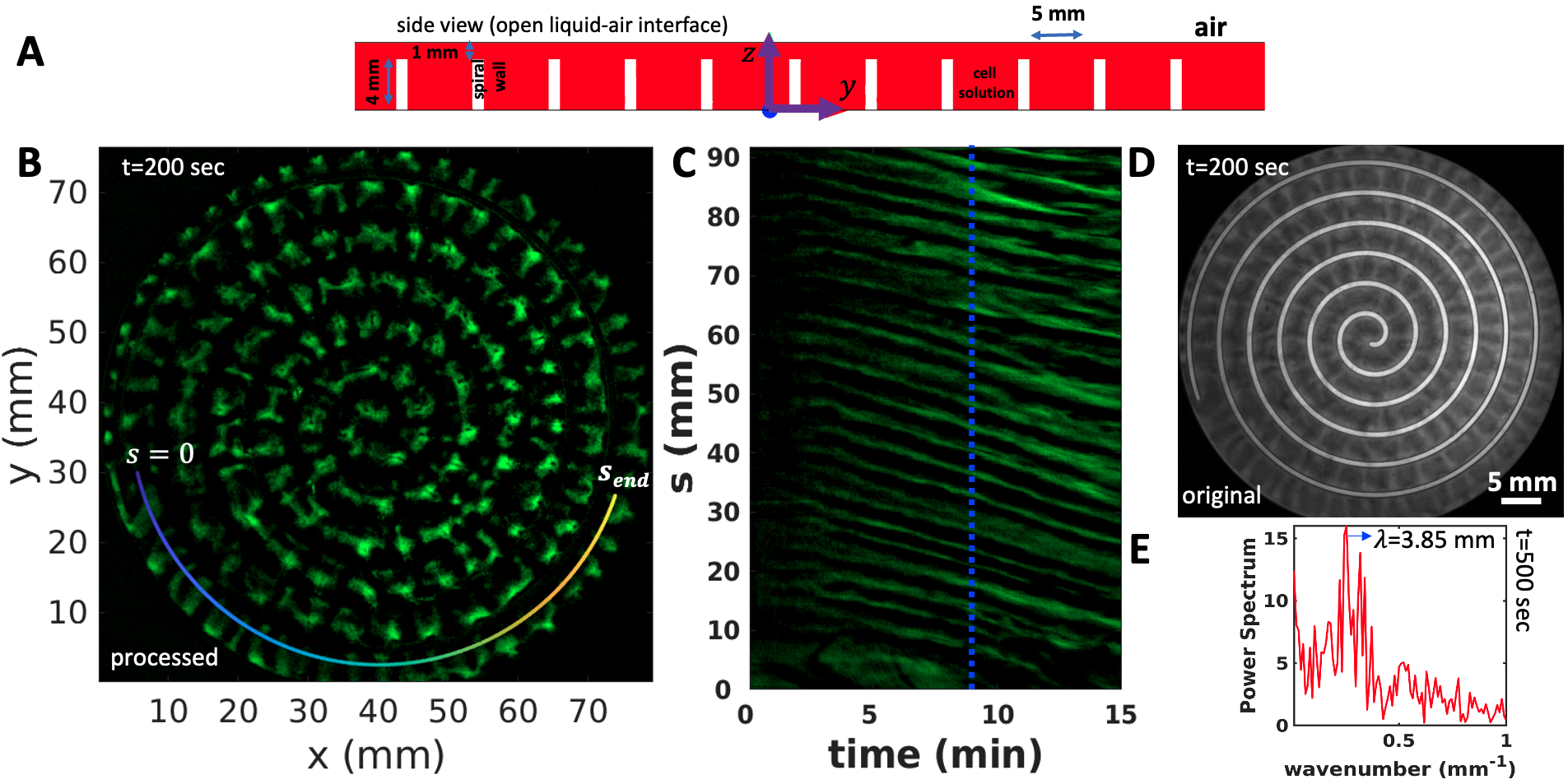}
	\includegraphics[width=\textwidth]{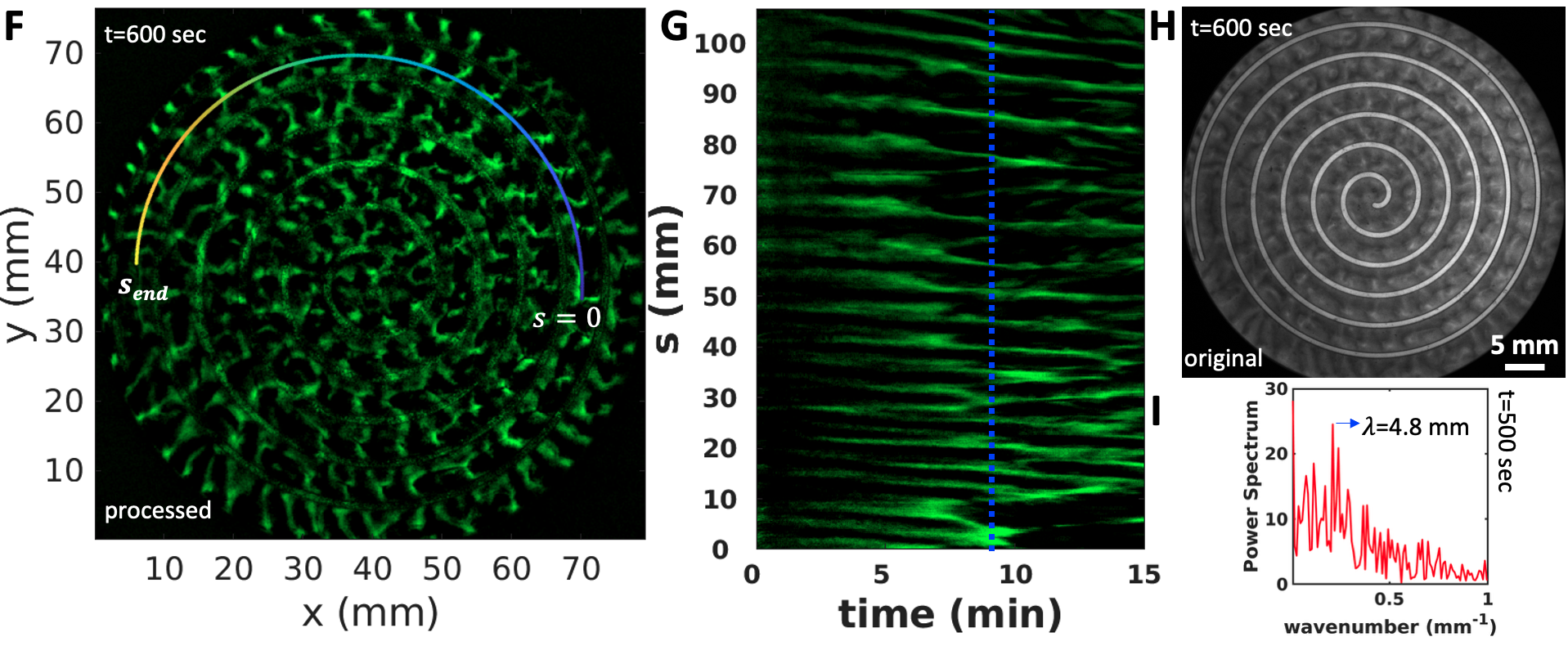}
	\includegraphics[width=\textwidth]{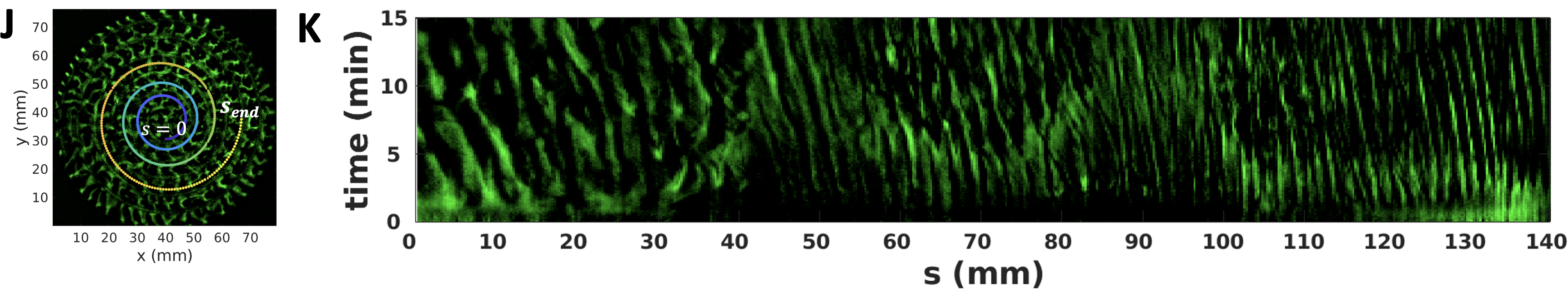}
	\end{subfigure}
	\hfill
	\begin{subfigure}[c]{0.49\textwidth}
		\centering
		\setlength\fboxsep{5pt}   
		\setlength\fboxrule{1.5pt}  
		\fcolorbox{green!70!black}{white}{
			\begin{minipage}{\linewidth}
				\centering
				\includegraphics[width=\linewidth]{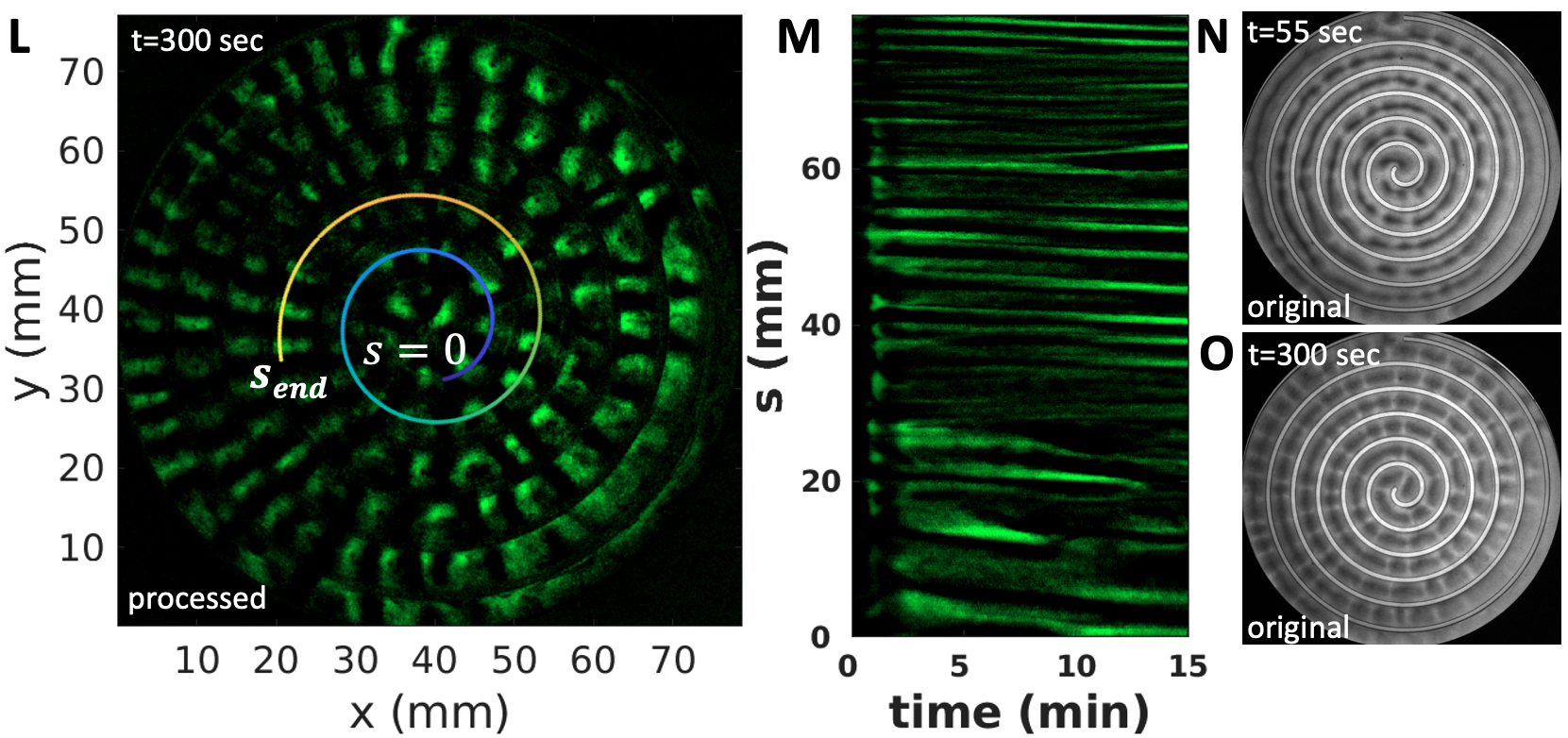}\\[1ex]
				\includegraphics[width=\linewidth]{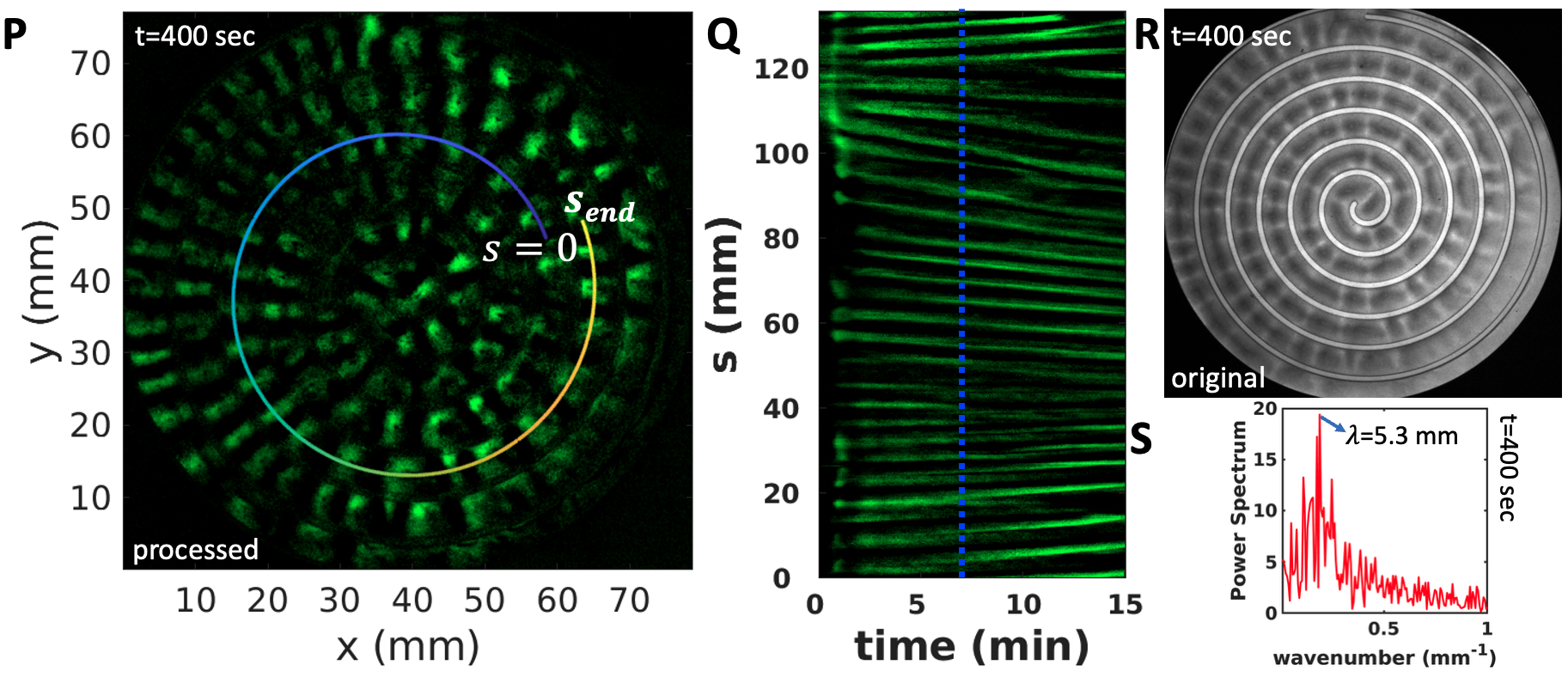}
			\end{minipage}%
		}
	\end{subfigure}
	\caption{\textcolor{black}{(A-K) Representative top-view images of bioconvection patterns formed by \textit{CR} cells confined within a spiral-shaped boundary with an open liquid–air interface show that, across nearly the entire setup, the patterns exhibit dominantly directed movement toward the spiral’s center. Panel (A) presents a side-view schematic of the experimental setup, indicating a fluid depth of 5\,mm, approximately 1\,mm higher than the height of the spiral boundary. Panels (B) and (D) show the early stages of pattern formation at the liquid–air interface, where cells accumulate due to negative gravitaxis, forming lattice-like clusters. As time progresses, these clusters become disordered, as seen in panels (F) and (H). Space–time plots in panels (C), (G), and (K), generated by stacking light intensity profiles along trajectories marked in panels (B), (F), and (J), reveal prominent inward-directed rotational dynamics across much of the domain. One-dimensional Fourier analysis along the dashed blue lines in panels (C) and (G) confirms a dominant pattern wavelength of approximately 4–5\,mm, as shown in panels (E) and (I); see Video 1. (L-S) An additional representative experiment under the same conditions as panel (A), except that pattern movement is inward near the spiral’s center (M) but proceeds either clockwise (CW) or counterclockwise (CCW) farther out (Q). Initially, cells accumulate at the liquid–air interface to form distinct spiral-shaped high-density regions (N), which then evolve into discrete, lattice-like structures (L, O, P and R); see Video 2.}}
	\label{fig:Spiral}
\end{figure*}
Two representative top-view examples of bioconvection patterns formed by \textit{CR} cells in a system with a spiral-shaped boundary and an open liquid–air interface are shown in Fig.~\ref{fig:Spiral}. In these experiments, the fluid depth is approximately 5\,mm, about 1\,mm higher than the height of the spiral boundary (Fig.~\ref{fig:Spiral}A). Starting from a homogeneous suspension, the cells swim upward due to negative gravitaxis, leading to a lower cell concentration directly above the spiral boundary relative to the deeper surrounding regions. This results in a characteristic spiral-shaped accumulation of cells at the liquid–air interface in the regions between the spiral walls (Fig.~\ref{fig:Spiral}N), which becomes susceptible to bioconvection instabilities. As the system evolves, this initial configuration fragments into smaller clusters, forming lattice-like patterns characterized by distinct dark spots—indicative of regions with high cell density—rendered as green in the post-processed images. These dense clusters are separated by brighter regions corresponding to areas of lower cell density, which appear dark in the post-processed images. Space–time plots generated along multiple trajectories across the setup reveal prominent large-scale rotational motion in the bioconvection patterns. Figure~\ref{fig:Spiral}A-K presents a representative experiment (observed in approximately 10\% of our total $N \approx 50$ experiments) in which the patterns exhibit dominant inward movement toward the spiral center across nearly the entire setup (Fig.~\ref{fig:Spiral}C, G, K). In contrast, Fig.~\ref{fig:Spiral}L-S illustrates the more commonly observed scenario, where directed inward rotational motion is primarily seen near the spiral core (Fig.~\ref{fig:Spiral}M), while regions farther from the center exhibit both clockwise (CW) and counterclockwise (CCW) rotational dynamics, with each region typically maintaining its initial direction of rotation over time (Fig.~\ref{fig:Spiral}Q). Fourier analysis performed along the trajectories marked by blue dashed lines in Fig.~\ref{fig:Spiral}C, G and Q reveals a characteristic pattern wavelength of approximately 4–5\,mm, consistently observed across various regions of the setup (Fig.~\ref{fig:Spiral}E, I, S).
\subsection{Oxygen deficiency triggers drastic pattern transitions  in  air impermeable setups}
\begin{figure}[!t]
	\begin{center}
		\includegraphics[width=\columnwidth]{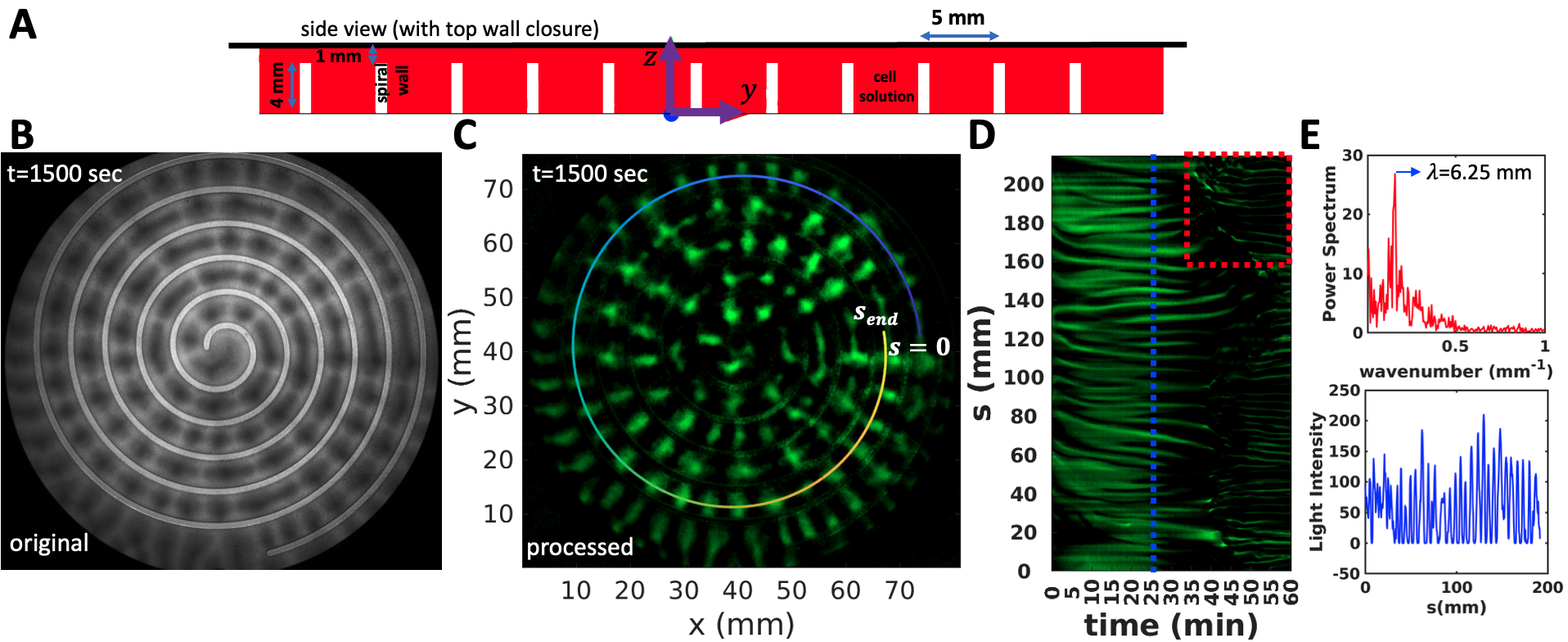}
		\includegraphics[width=\columnwidth]{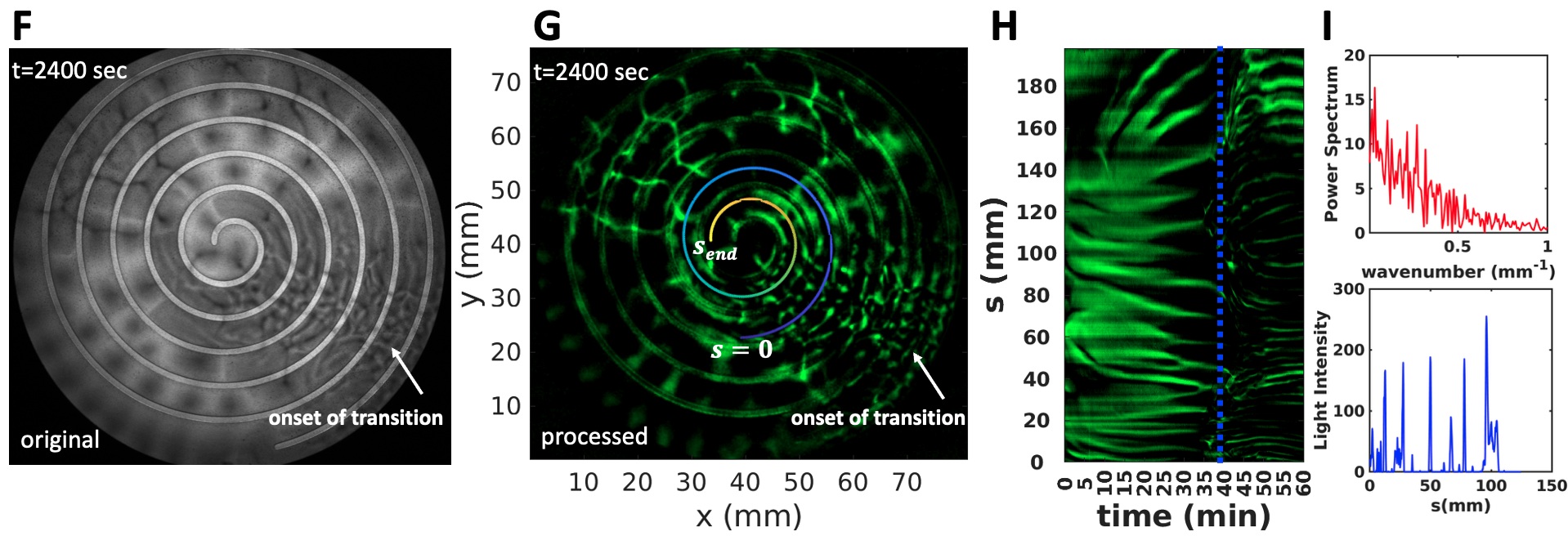}
		\includegraphics[width=\columnwidth]{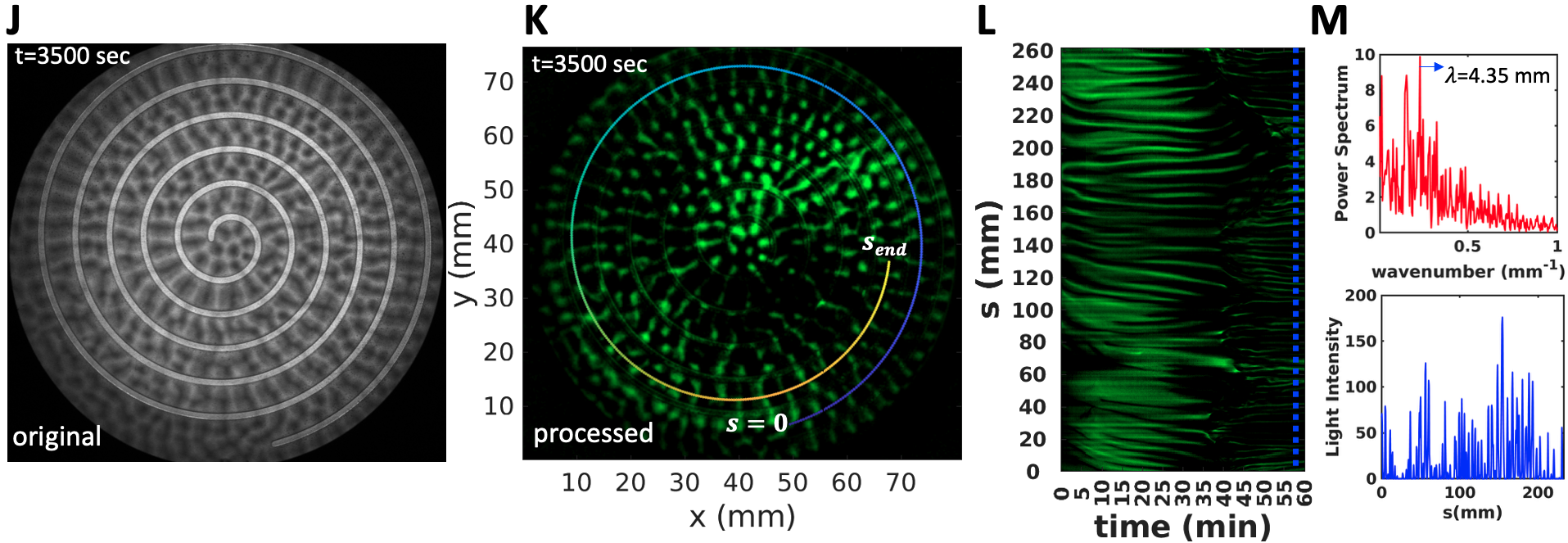}
		\caption{Evolution of bioconvection patterns under confined air-impermeable conditions with a closed top boundary. (A) Experimental setup with a transparent PMMA wall placed above the spiral structure, maintaining a 1\,mm gap. (B–D) Initial lattice-like arrangement of high-density cell clusters (visualized as green spots in post-processed images) separated by lower-density regions. (F–H) Transition phase characterized by fragmentation and the formation of a disorganized, branched cellular network. (J–L) Reorganization into a new, stable pattern with a reduced wavelength. (E, I, M) One-dimensional Fourier transforms of light intensity profiles, extracted along the blue dashed lines in panels D, H, and L, reveal a reduction in the dominant wavelength from approximately 6\,mm to 4\,mm. See also supplementary Video 3 for the full dynamic evolution.} 
		\label{fig:Spiral_Closed}
	\end{center}
\end{figure}
We modified the boundary conditions by placing a solid transparent PMMA wall above the spiral structure, maintaining a 1\,mm gap between the spiral wall and the top boundary. The PMMA-based chamber was completely filled with the cell suspension, ensuring the elimination of any air gaps or bubbles (Fig.~\ref{fig:Spiral_Closed}A). Under these confined, air-impermeable conditions, we observed a pronounced transition in the bioconvection patterns over time, which we attribute to oxygen depletion. To characterize this transition, we carried out frame-by-frame analysis, constructed detailed space-time plots, and applied one-dimensional Fourier transforms (1D-FT) to quantify changes in the pattern wavelengths. We measured the wavelengths at three critical stages: before the transition, at its onset, and after the transition. Long-duration recordings (approximately two hours) revealed a significant reduction in the characteristic wavelength, accompanying a dramatic reorganization of the bioconvection patterns.

Similar to the setup with an open liquid-air interface, the bioconvection patterns initially emerged from a homogeneous suspension as localized accumulations of cells arranged in a lattice-like structure. These patterns were characterized by distinct dark regions, corresponding to areas of high cell density (visualized as green spots in post-processed images), separated by brighter regions of lower cell concentration (appearing dark in processed images); see Fig.~\ref{fig:Spiral_Closed}B–C and Video 3. This lattice arrangement persisted for approximately 30–40 minutes before undergoing a drastic transition (Fig.~\ref{fig:Spiral_Closed}F-M). The transition was marked by a temporary weakening or disappearance of the regular pattern, followed by the formation of a disorganized, branched network of cells at random locations (Fig.~\ref{fig:Spiral_Closed}F–I), which rapidly percolated throughout the entire suspension, resulting in a brief intermediate phase of disorganization.  Following this phase, the cells spontaneously reorganized into a new, stable, and more regular pattern with a smaller characteristic wavelength (Fig.~\ref{fig:Spiral_Closed}J–M), which remained steady for over an hour as long as the cells retained their motility. Note that the upper portion of the transitional zone—highlighted by a red box in Fig.~\ref{fig:Spiral_Closed}D—exhibits a temporal delay, revealing that the transition event propagates gradually across the suspension. To quantify the associated changes in the pattern wavelength, we applied one-dimensional Fourier transforms to the light intensity profiles extracted from the space-time plots at the selected time points. \textcolor{black}{The results presented in Fig.~\ref{fig:Spiral_Closed}E and M reveal a reduction in the dominant wavelength from roughly $6\,\mathrm{mm}$ to $4\,\mathrm{mm}$ during the transition. This $\sim2\,\mathrm{mm}$ decrease was consistent across all experiments exhibiting a pattern transition.}
\begin{figure}[!t]
	\begin{center}
		\includegraphics[width=\columnwidth]{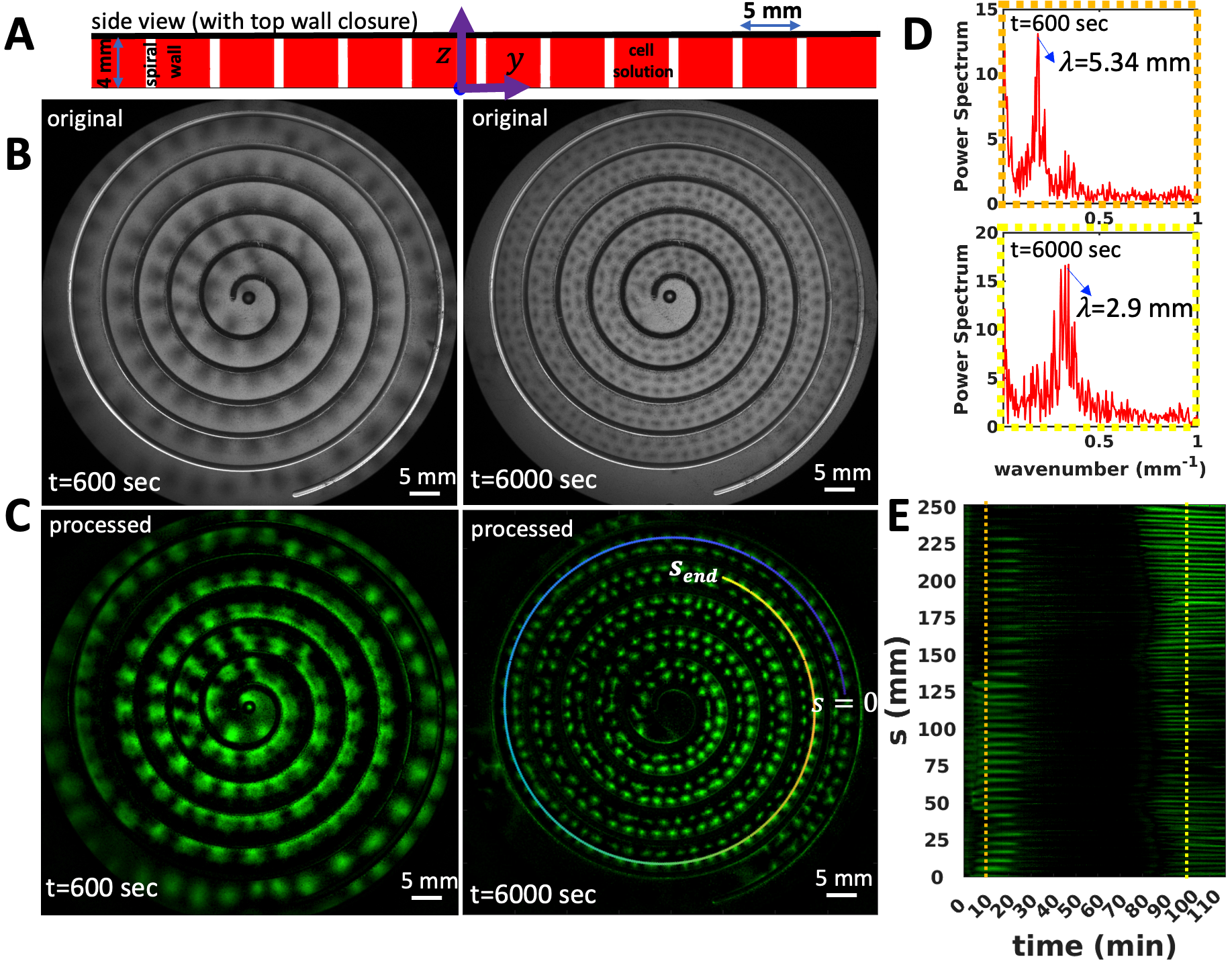}
		\caption{Bioconvection pattern evolution under fully confined air-impermeable conditions with no fluid gap. (A) Experimental setup where the top PMMA wall directly contacts the spiral-shaped boundary, eliminating the 1\,mm fluid gap present in the previous configuration. (B–D) Sequential top-view images showing the disappearance or significant weakening of the initial bioconvection pattern, followed by reorganization into a new, stable pattern with a smaller characteristic wavelength. (E) Space–time plot illustrating the transition, highlighting two distinct regions with different stripe spacings, separated by a prolonged intermediate phase of disorganization. The extended transition time is attributed to weakened cell-to-cell hydrodynamic interactions due to the absence of a fluid gap. Fourier analysis confirms a decrease in the dominant wavelength from approximately 5\,mm to 3\,mm (panel D); see Video 4 for the full dynamic evolution.} 
		\label{fig:PeriodDoubling}
	\end{center}
\end{figure}
\begin{figure}[!t]
	\begin{center}
		\includegraphics[width=\columnwidth]{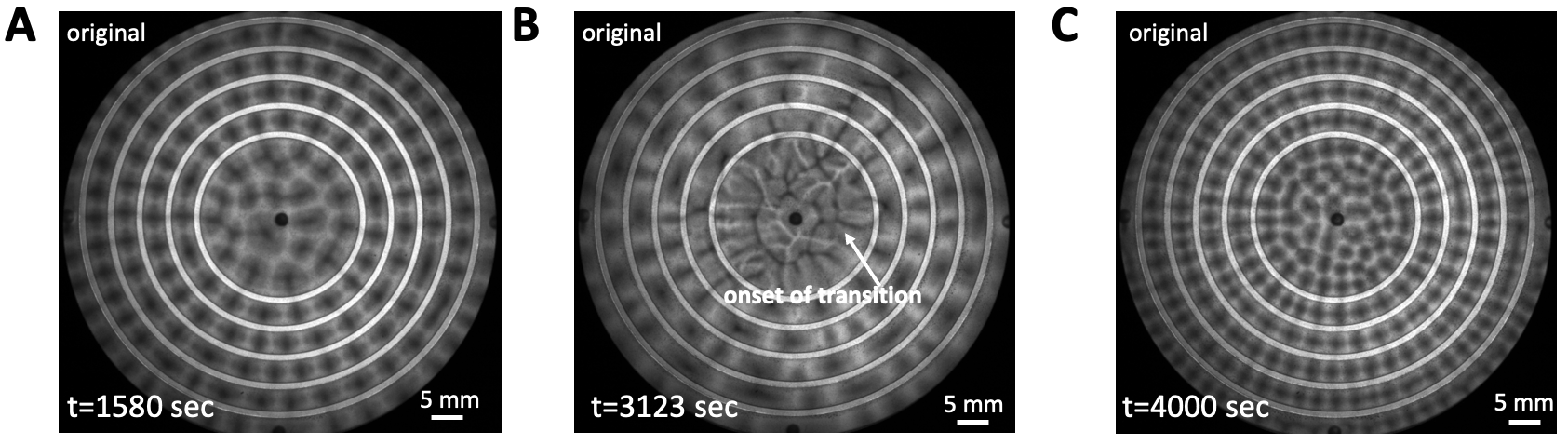}
		\includegraphics[width=\columnwidth]{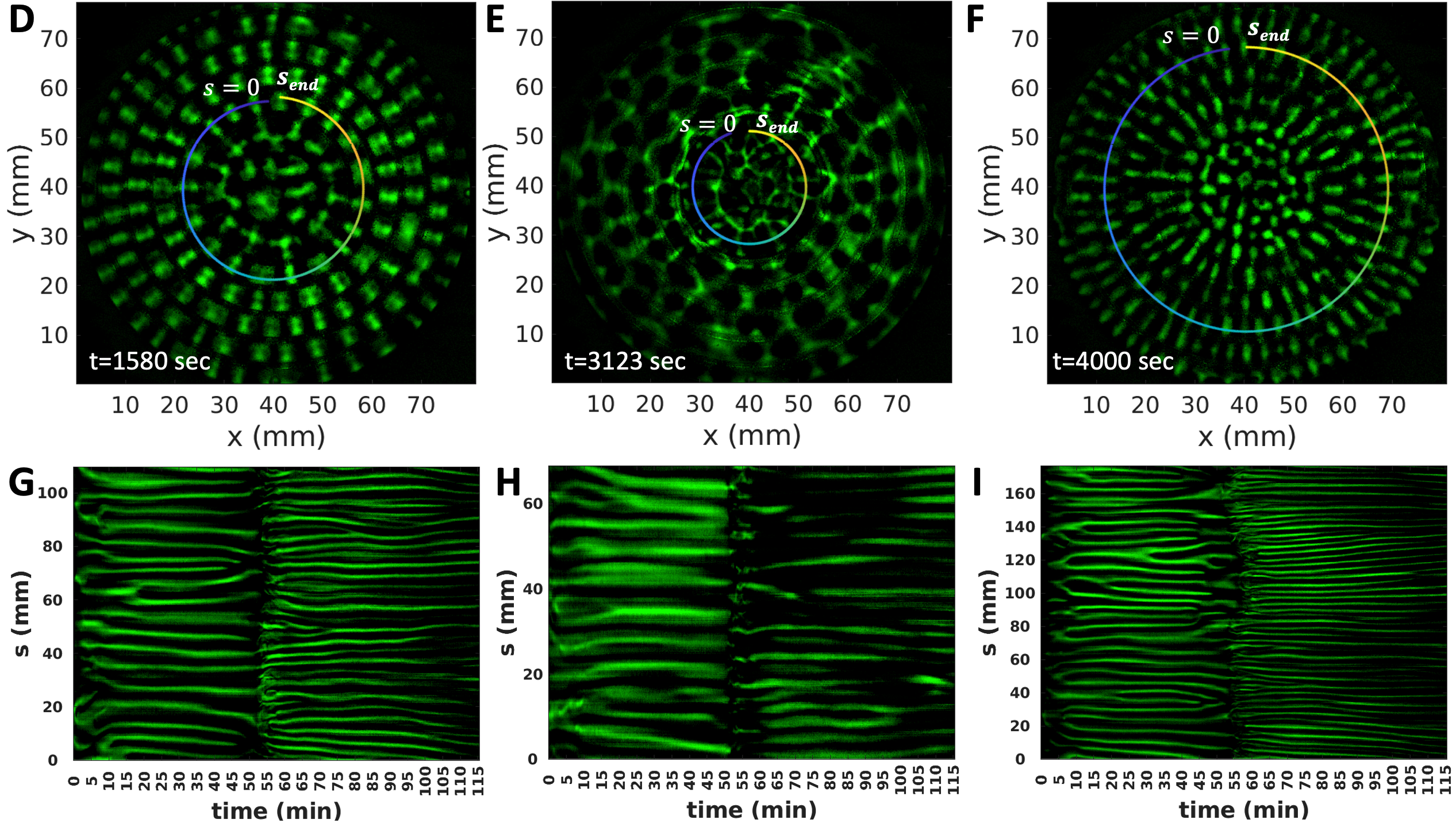}
		\caption{Bioconvection pattern evolution under confined air-impermeable conditions with a 1\,mm fluid gap (as sketched in Fig.~\ref{fig:Spiral_Closed}A) in a PMMA-made setup containing circular obstacles. (A–C) Sequential top-view images showing the reorganization of bioconvection patterns, with a new pattern featuring a smaller characteristic wavelength emerging after approximately 50 minutes. (D–F) Corresponding processed images of the snapshots shown in panels A–C. (G-I) Space–time plots along the trajectories indicated in panels D–F, revealing the transition, characterized by two distinct regions with different stripe spacings; see Video 5.} 
		\label{fig:PeriodDoubling_Circular}
	\end{center}
\end{figure}

Building on these observations, we conducted additional experiments under confined conditions, modifying the setup so that the top PMMA wall was in direct contact with the upper edge of the spiral-shaped boundary, thereby eliminating the 1\,mm fluid gap present in the previous configuration (Fig.~\ref{fig:PeriodDoubling}A). Sequential top-view images captured at different stages of bioconvection pattern evolution are shown in Fig.~\ref{fig:PeriodDoubling}B--C. Under these fully confined air-impermeable conditions, the intermediate phase---during which the patterns disappeared or substantially weakened---was significantly prolonged, lasting approximately 50 minutes before the system reorganized into a new, stable configuration with a shorter characteristic wavelength (Fig.~\ref{fig:PeriodDoubling}D). This prolonged transition and subsequent reorganization are clearly illustrated in the space--time plot (Fig.~\ref{fig:PeriodDoubling}E), which reveals two distinct regions with different stripe spacings corresponding to pre- and post-transition states, separated by an extended intermediate phase characterized by weak or absent pattern formation. We attribute this delay to reduced fluid-mediated hydrodynamic interaction between cells due to the absence of the 1\,mm fluid gap, which in the previous configuration facilitated more efficient collective reorganization. Despite the slower dynamics, Fourier analysis confirms a reduction in the dominant pattern wavelength from approximately 5\,mm to 3\,mm following the transition.
To further confirm that oxygen depletion is the underlying cause of this transition, we performed a control experiment in which the entire chamber, including the top boundary, was fabricated from air-permeable PDMS. Under otherwise identical conditions, no pattern transitions were observed (Fig.~\ref{fig:All_PDMS}), strongly supporting the conclusion that restricted oxygen exchange in PMMA-based confined geometries is the key factor driving the observed drastic pattern reorganization.

To assess the generality of these findings, we extended our experiments to systems incorporating alternative internal geometries. Specifically, we replaced the spiral boundary with an array of circular obstacles while preserving the same confined conditions used in Fig.~\ref{fig:Spiral_Closed}, including the 1\,mm fluid gap. As illustrated in Fig.~\ref{fig:PeriodDoubling_Circular}, similar transitions in bioconvection patterns were observed, with the structures reorganizing into a new configuration featuring a reduced characteristic wavelength after roughly 50 minutes. These results demonstrate that the key driver of the transition is the oxygen-depletion itself rather than the specific geometry of the internal boundary. Thus, the striking pattern transitions reported here emerge as robust features of confined bioconvection within air-impermeable boundaries, occurring regardless of whether the internal geometry is spiral, circular, or of another structural configuration.
\subsection{Simulations}
\begin{figure}[!t]
	\begin{center}
		\includegraphics[width=0.9\columnwidth]{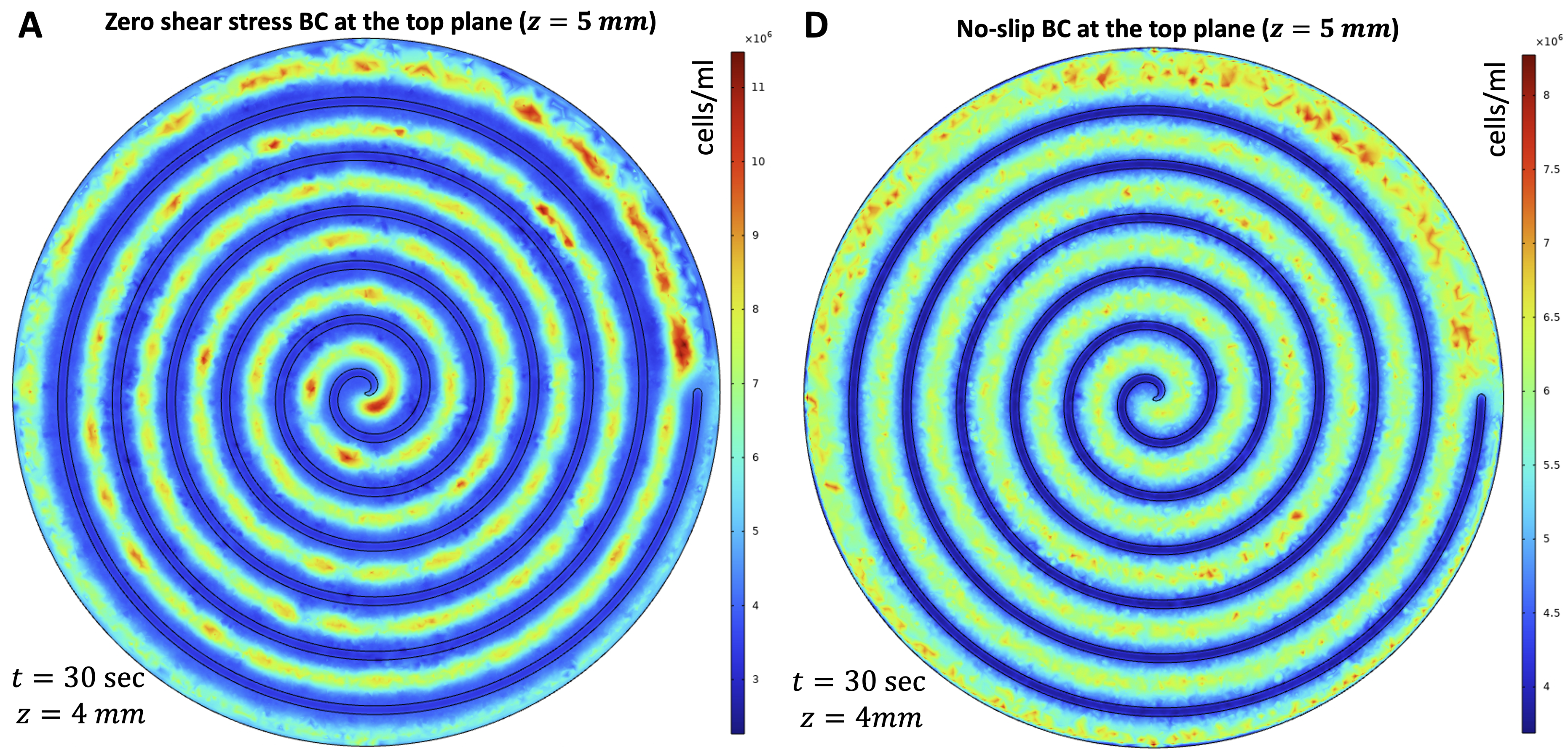}
		\includegraphics[width=0.9\columnwidth]{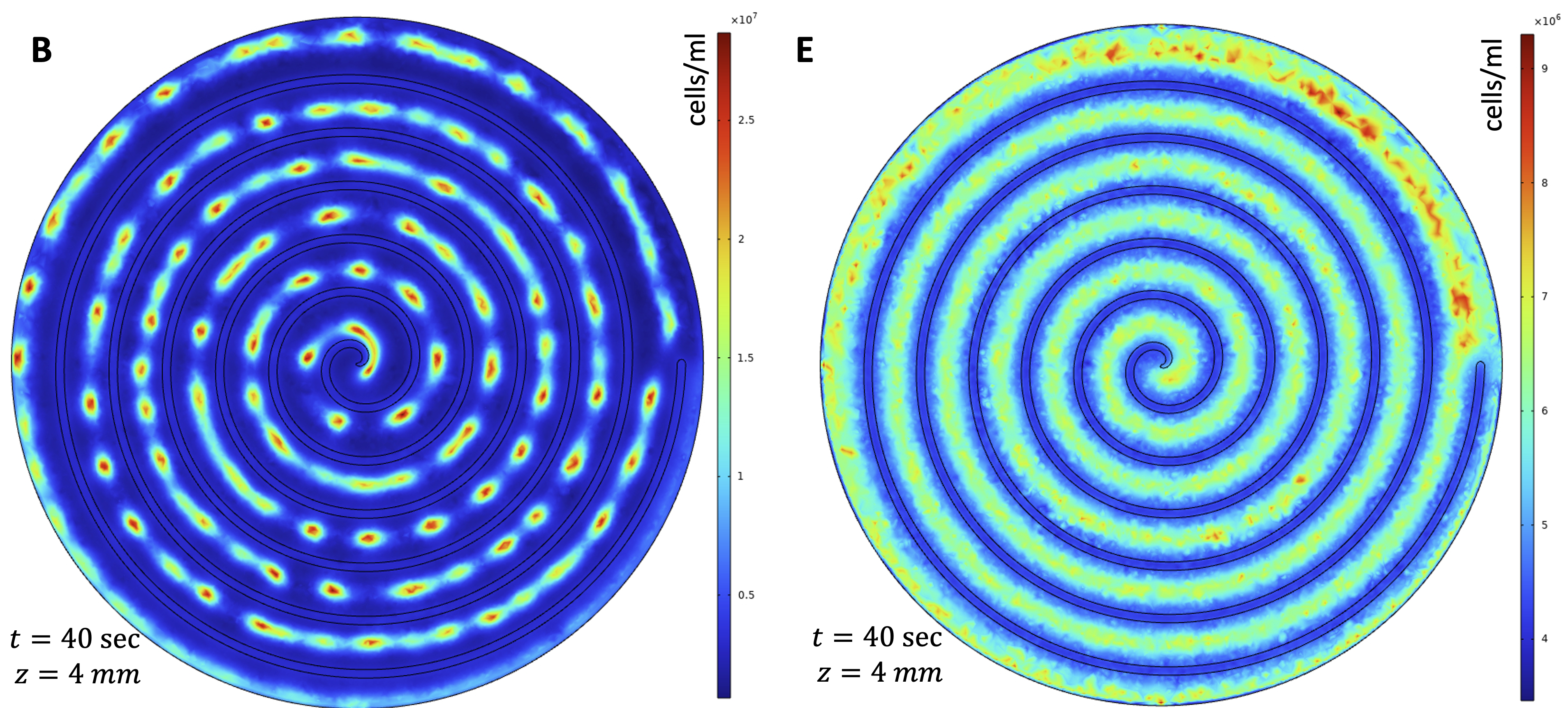}
		\includegraphics[width=0.9\columnwidth]{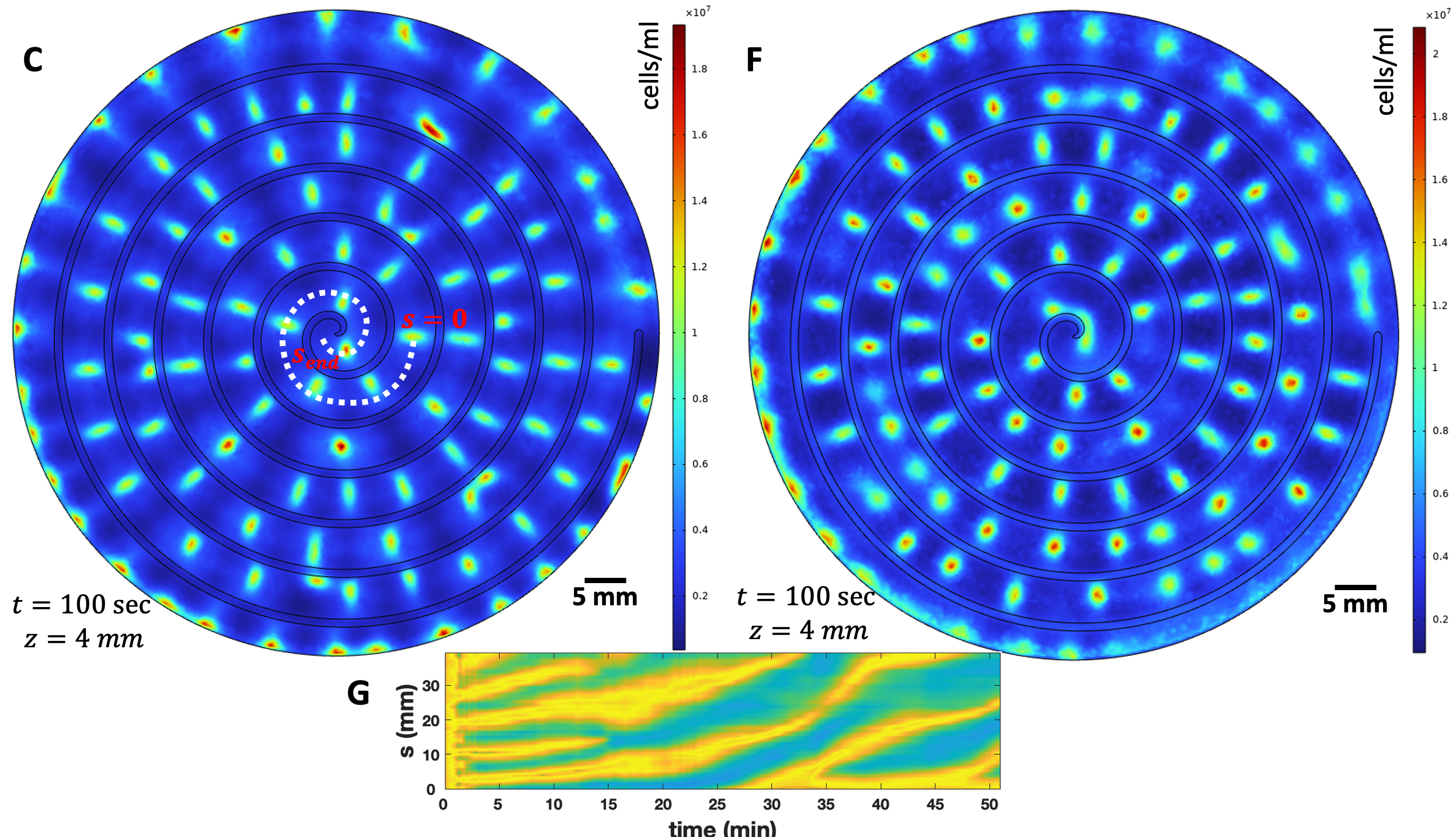}
		\caption{Top-view snapshots of cell density from 3D simulations of bioconvection patterns above the spiral-shaped boundary at $z = 4$\,mm. Either a zero-shear-stress condition (representing a free liquid-air interface) or a no-slip condition (representing a rigid top wall) is applied at the upper boundary (at $z=5$\,mm). Initially, \textit{CR} cells swim upward (negative gravitaxis) and accumulate near the surface, forming a distinct spiral-shaped cell density pattern (A, D). This structure becomes dynamically unstable, fragmenting into smaller, high-density clusters that evolve into down-welling plumes (see also Fig.~\ref{fig:Flow}). Notably, fragmented clusters, initially aligned parallel to the spiral walls (B), progressively reorient perpendicular to the boundaries over time (C, F). A comparison between panels B and E shows that fragmentation of the spiral-shaped accumulation occurs earlier in the system with the zero-shear-stress boundary condition. G) The space–time plot along the spiral trajectory in panel C, constructed from long-term simulation results, reveals the inward movement of plumes toward the spiral center, in good agreement with experimental observations. The color map has been slightly adjusted to enhance visualization; see also Videos 6-8.} 
		\label{fig:Simulations_Z4mm}
	\end{center}
\end{figure}
We investigate three-dimensional bioconvection patterns within a shallow suspension of cells contained in a cylindrical chamber. The simulated geometry closely matches our experimental setup and consists of a rigid bottom surface, rigid cylindrical sidewalls, and a rigid spiral-shaped internal wall immersed in the cell suspension. The fluid layer extends approximately 1\,mm above the spiral wall, which has a height of about 4\,mm. We apply no-slip boundary conditions at all solid surfaces. At the upper boundary, we impose either a zero-shear-stress boundary condition to represent a free liquid-air interface or a no-slip condition to mimic a rigid top boundary in confined geometry experiments. Additionally, we ensure no flux of cells occurs through the top, bottom, or side boundaries.

\begin{figure}[!t]
	\begin{center}
		\includegraphics[width=\columnwidth]{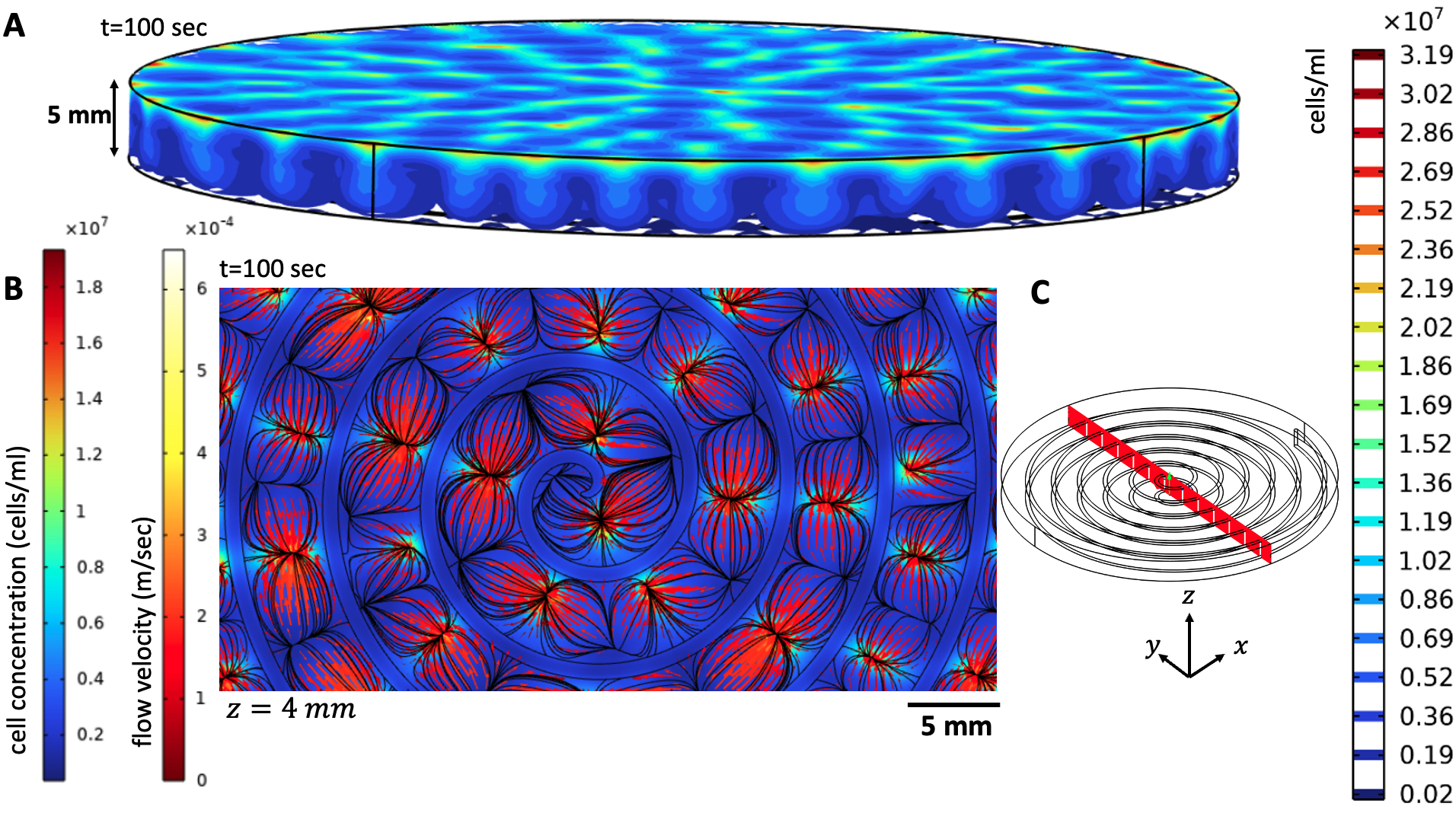}
		\caption{(A) Three-dimensional visualization of emerging bioconvection patterns within the setup, represented by iso-surface concentration contours. (B) Zoomed-in top view of the cell density color map and surface flow field with streamlines at $z=4$\,mm, highlighting the formation of downwelling plumes. (C) Vertical cut-plane used to extract the density profiles and flow fields shown in Fig.~\ref{fig:Flow}; see also Video 11 and 12.} 
		\label{fig:3D}
	\end{center}
\end{figure}

\begin{figure}[!t]
	\begin{center}
		\includegraphics[width=\columnwidth]{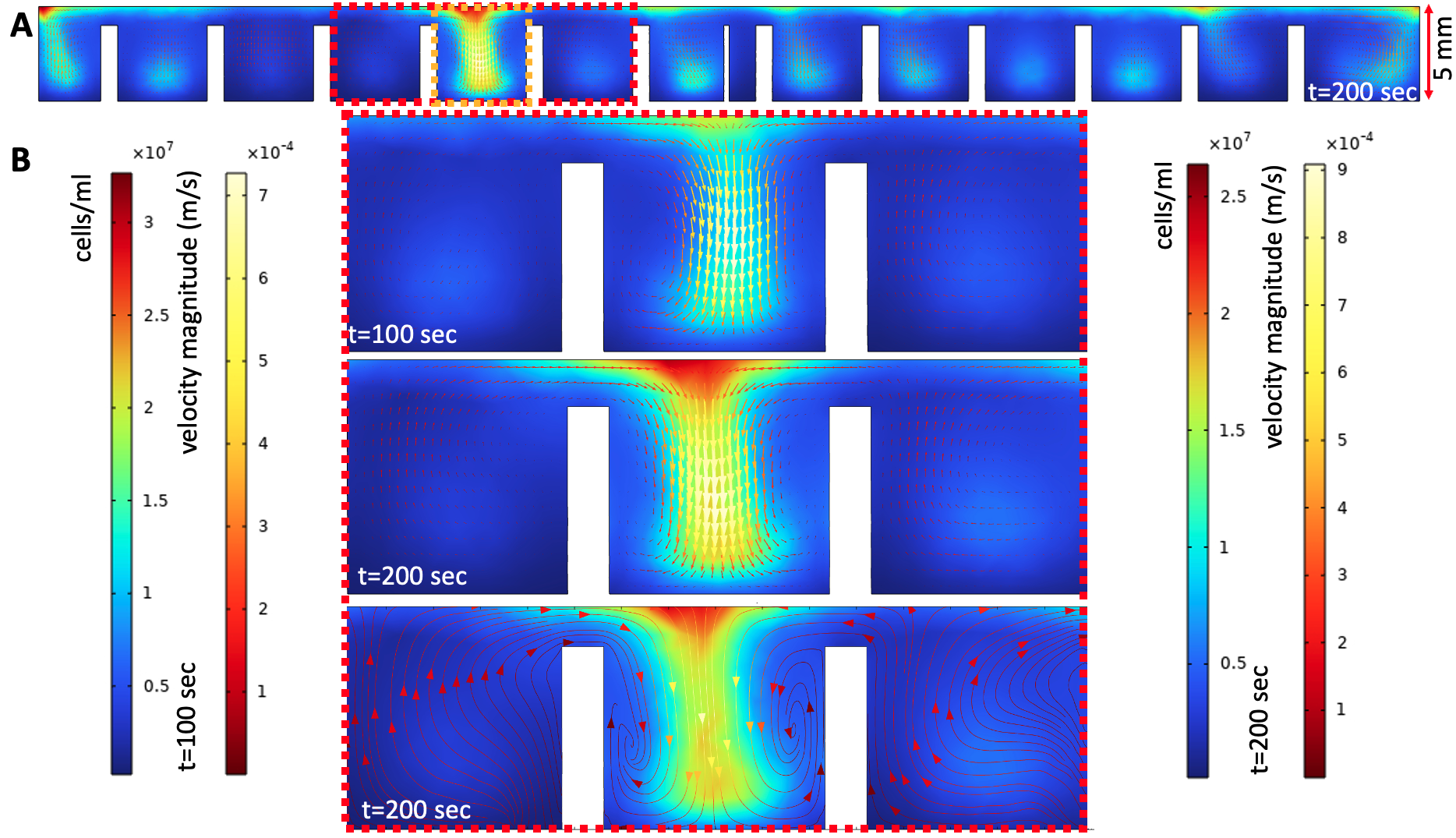}
		\includegraphics[width=\columnwidth]{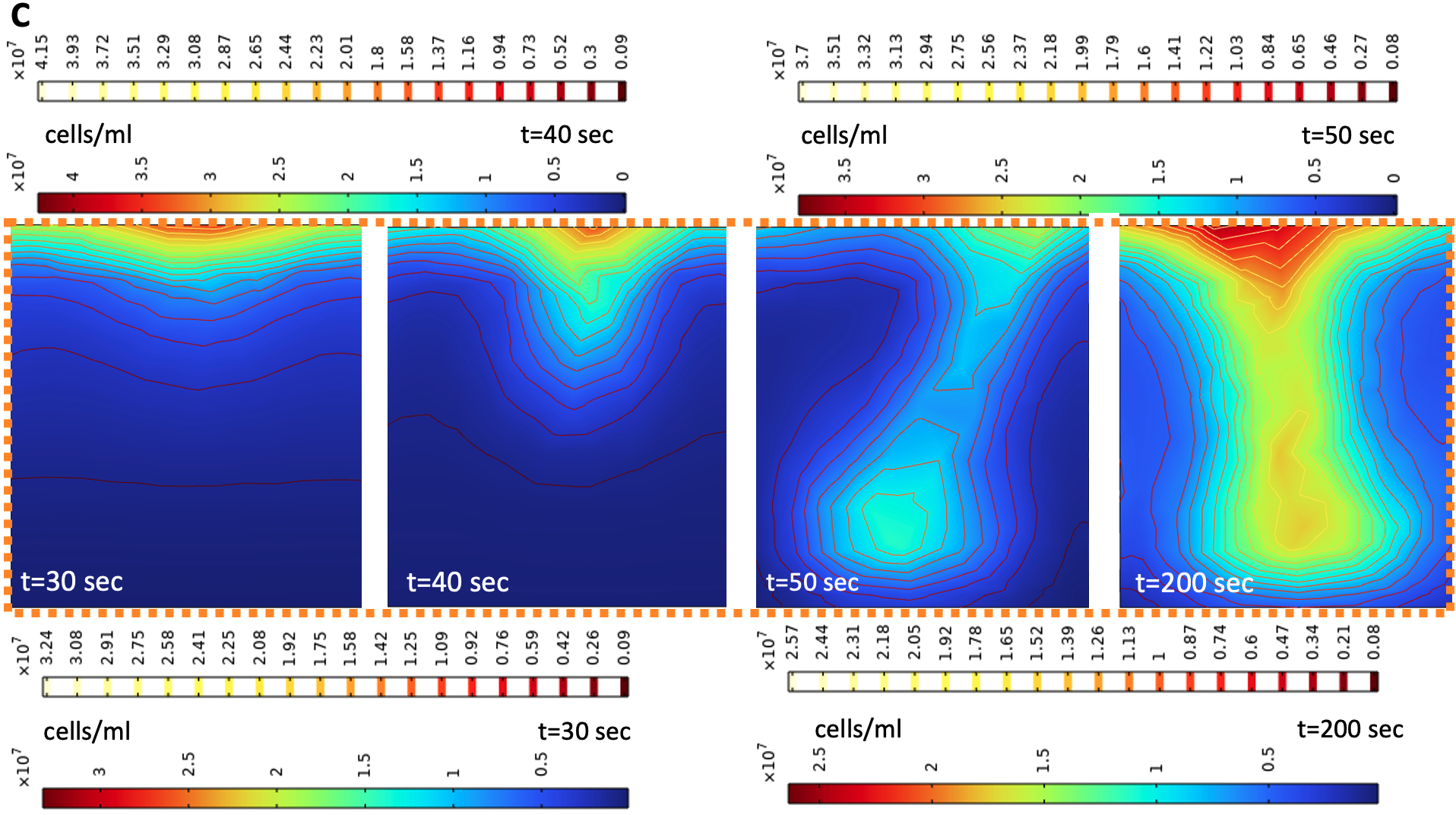}
		\caption{\textcolor{black}{Cell density distribution and corresponding flow field with streamlines extracted along a vertical cut plane passing through the center of the spiral-shaped boundary (as defined in Fig.~\ref{fig:3D}C). Panel B shows a zoomed-in view of the region highlighted by the dashed red box in panel A, revealing the formation of large-scale vortices and strong downward plumes, with fluid velocities reaching approximately 1\,mm/s—nearly an order of magnitude greater than the typical swimming speed of individual \textit{CR} cells. The dashed orange box in panel A indicates the region further magnified in panel C.  (C) Cell density distribution and corresponding iso-density contour lines during the formation of downwelling plumes, showing zoomed-in views of the region highlighted by the orange box in panel A at successive time points; see Videos 13 and 14.}} 
		\label{fig:Flow}
	\end{center}
\end{figure}

Figure~\ref{fig:Simulations_Z4mm} shows a representative top-view of bioconvection patterns that form directly above the spiral-shaped boundary at a height of $z = 4$\,mm, where $z = 0$ corresponds to the bottom of the chamber and $z = 5$\,mm marks the liquid–air interface. At early times, \textit{CR} cells exhibit negative gravitaxis, swimming upward against gravity and accumulating near the free liquid-air interface (see also the side view in Fig.~\ref{fig:Flow} and the three-dimensional density profile in Fig.~\ref{fig:3D}A). Owing to the shallower fluid layer directly above the spiral walls—approximately 1\,mm compared to 5\,mm in surrounding regions—the cell concentration at the surface is initially lower above the spiral boundaries than in adjacent areas. This variation in fluid depth seeds the formation of a distinct spiral-shaped accumulation pattern between the spiral walls, as shown in Fig.~\ref{fig:Simulations_Z4mm}A, D and Fig.~\ref{fig:Simulations_Z5mm}A, D, in agreement with experimental observations (Fig.~\ref{fig:Spiral}N). However, this spiral-shaped accumulation is dynamically unstable due to density inversion, as \textit{CR} cells are approximately 5\% denser than water. As a result, the initially continuous spiral structure fragments into smaller, high-density clusters (Figs.~\ref{fig:Simulations_Z4mm}B, E), which subsequently evolve into distinct downwelling plume structures (Fig.~\ref{fig:Flow}). The formation of these downwelling plumes is accompanied by the emergence of surface flows, as illustrated in Fig.~\ref{fig:3D}B. Notably, the fragmented clusters at $z=4$\,mm, which initially align parallel to the spiral walls, progressively reorient perpendicular to the boundaries over time, as depicted in Figs.~\ref{fig:Simulations_Z4mm}C and F.

This reorganization of cell clusters near the surface raises the question of how the underlying bioconvection structures develop throughout the depth of the suspension. To investigate this, we examine the cell density distribution at the bottom of the setup. Figure~\ref{fig:Simulations_Z0mm} shows the cell density color map at $z = 0$. Interestingly, comparing the snapshots at $t = 100$\,sec with the corresponding near-surface patterns in Fig.~\ref{fig:Simulations_Z4mm} reveals that not only the cell concentration is approximately two times lower at the bottom, but also that the cell clusters remain more diffuse and are not oriented perpendicular to the spiral boundary, in contrast to the reorganization observed near the surface in Figs.~\ref{fig:Simulations_Z4mm}C and F.

In addition to characterizing the evolving cell density patterns, we investigated the large-scale coherent flow fields generated as downwelling plumes developed within the suspension. To quantify the fluid dynamics associated with bioconvection, velocity fields were extracted along a vertical cut-plane intersecting the center of the spiral-shaped boundary, as defined in Fig.~\ref{fig:3D}C. Figure~\ref{fig:Flow} presents the cell density distribution together with the corresponding velocity field and overlaid streamlines in this plane. To further illustrate the onset of flow instabilities, Fig.~\ref{fig:Flow}C displays iso-density contour lines highlighting the emergence and structure of the developing plumes. Consistent with previous observations reported in the literature~\cite{bees2020advances}, we find the formation of vortices featuring strong downward flows concentrated within the plumes, with velocities reaching approximately 1\,mm/s—nearly an order of magnitude greater than the typical swimming speed of individual \textit{CR} cells. These results highlight how collective cell motion dramatically enhances fluid transport far beyond the capabilities of isolated swimmers.
\section{Discussion and Conclusions}
Our experimental observations reveal striking transitions in bioconvection patterns driven by oxygen depletion, highlighting the critical interplay between boundary conditions, oxygen availability, and microbial collective behavior. Under confined conditions—achieved by placing a transparent PMMA top wall 1\,mm above the spiral-shaped boundary—we initially observed stable lattice-like bioconvection patterns, closely resembling those seen in systems with an open liquid-air interface. As the cells gradually consumed the dissolved oxygen, these regular patterns weakened significantly or temporarily disappeared, followed by a spontaneous reorganization into a new, stable arrangement with a markedly smaller wavelength. The transition was initiated at random locations and propagated across the suspension, indicating that localized oxygen depletion progressively modulated cell motility and metabolic activity, thereby driving the onset of collective reorganization. When confinement was further intensified—by eliminating the fluid gap entirely—we observed even more pronounced and prolonged transitions. Under these fully confined conditions, the intermediate phase, during which the bioconvection patterns disappeared, extended to approximately 45 minutes before the system spontaneously reorganized into a new configuration with a reduced wavelength. Reorganization typically began at random locations and gradually percolated throughout the entire suspension. Although oxygen levels are ultimately expected to deplete uniformly throughout the suspension, early-stage spatial variations in cell concentration leads to heterogeneous local depletion rates, causing pattern reorganization to initiate at random locations across the system. Collectively, these results demonstrate that subtle changes in confinement geometry and oxygen availability can profoundly influence the dynamics, timescales, and structural evolution of bioconvection patterns, offering promising strategies for manipulating microbial collective behavior through environmental and geometric control.

Our findings are consistent with the recent study by Fragkopoulos \textit{et al.}~\cite{fragkopoulos2025metabolic}, who demonstrated that under anaerobic conditions, confined suspensions of photosynthetically active microorganisms exhibit reversibly switchable coherent flows. In their experiments, lowering the red-light intensity under anaerobic conditions—thereby reducing photosynthetic activity—led to a marked decrease in cell swimming speeds (from approximately 100\,$\mu$m\,s$^{-1}$ to 40\,$\mu$m\,s$^{-1}$) and initiated the emergence of spatially regular, temporally stationary bioconvective plumes within an otherwise stable suspension exhibiting an inverted sedimentation profile. While Fragkopoulos \textit{et al.} systematically varied the light intensity to modulate metabolic activity, in our experiments the red-light intensity was held constant at approximately $5\times10^{17}$ photons \,m$^{-2}$\,s$^{-1}$ throughout. Instead, oxygen availability was modulated purely through physical confinement. We propose that confinement-induced oxygen depletion under low-light conditions triggers a metabolic shift from aerobic to anaerobic respiration, which in turn drives the observed transition of bioconvection patterns. Thus, both work independently support the conclusion that metabolic activity and oxygen dynamics are key regulators of collective behavior in motile microbial suspensions and further emphasizes that under low-light conditions geometric confinement alone, by limiting oxygen replenishment, can drive dramatic reorganizations of bioconvection patterns even in otherwise constant environmental conditions.

Furthermore, an earlier study by Kage \textit{et al.}~\cite{kage2013drastic} reported drastic reorganization of bioconvection patterns in confined \textit{CR} suspensions under red-light illumination, which they primarily attributed to gyrotactic reorientation—namely, changes in the balance between gravitational and viscous torques experienced by the cells. In that study, the authors also observed that pattern wavelength shifts correlated with illumination intensity, noting that red light predominantly influenced the steady-state patterns after the transition, rather than affecting the initial onset or the pre-transition steady states. They proposed that these changes could be linked to alterations in cellular photosynthetic activity. In connection with these findings, our results—obtained under constant red-light illumination—highlight that confinement, and the associated oxygen depletion, likely inducing a shift from aerobic to anaerobic metabolism, serves as a critical driver of the observed pattern transitions. Notably, in our experiments, no pattern transitions were observed under open liquid–air interface conditions, nor in control experiments where the entire setup—including the top boundary—was fabricated from oxygen-permeable PDMS (Fig.~\ref{fig:All_PDMS}). In contrast, striking pattern reorganizations consistently emerged under confined PMMA-based configurations with restricted oxygen replenishment, underscoring the critical role of oxygen availability in driving the observed transitions.

To better understand the mechanisms underlying the pattern transitions observed in our experiments, we performed numerical simulations based on standard continuum models of bioconvection. These models, which couple the incompressible Navier–Stokes equations with cell conservation and gyrotactic reorientation~\cite{pedley1992hydrodynamic,bees1999non}, have been widely used to describe bioconvection instabilities. In these frameworks, the onset of bioconvection is predicted when the Rayleigh number exceeds a critical threshold, with characteristic pattern wavelengths determined by the fastest-growing unstable modes. However, it is important to recognize the inherent limitations of these classical models. They typically assume constant swimming speeds and negligible cell–cell interactions under the dilute-suspension approximation. As noted previously~\cite{bees1999non,hill2005bioconvection}, in the nonlinear stages of bioconvection, local cell concentrations within dense, downwelling plumes can significantly exceed the initial mean value, challenging the validity of the dilute-suspension assumption and potentially altering hydrodynamic and collective behaviors. In our confined system, the observed pattern transitions—characterized by the temporary disappearance and subsequent reorganization into smaller-wavelength structures—cannot be explained solely by these classical bioconvection models. Even when our numerical simulations incorporated no-slip boundary conditions to emulate experimental confinement, the standard framework was only able to reproduce the initial formation and fragmentation of spiral-shaped cell accumulations into downwelling plumes; it failed to capture the striking reorganizations into new, stable patterns observed experimentally. This discrepancy highlights the need to extend standard bioconvection theories to account for additional factors, notably oxygen dynamics. Our experiments strongly suggest that oxygen depletion—and the associated transition from aerobic to anaerobic metabolism, which was shown to modulate microbial motility in Ref.~\cite{fragkopoulos2025metabolic}—plays a critical role in driving these transitions. 

{\color{black}Inspired by models of oxytactic bacterial suspensions and photosynthetic microbes (e.g., Refs.~\cite{hill2005bioconvection,keller1971traveling,hopkins2002computational,desai2017modeling,hillesdon1996bioconvection,fragkopoulos2021self}), we introduce an explicit oxygen concentration field \(C_{\mathrm{oxy}}(\mathbf{x},t)\) and couple it to the bioconvection flow \(\mathbf{U}(\mathbf{x},t)\) and local cell concentration \(c(\mathbf{x},t)\) via the reaction–diffusion–advection equation:
\begin{equation}
	\frac{\partial C_{\mathrm{oxy}}}{\partial t}
	+ \mathbf{U}\cdot\nabla C_{\mathrm{oxy}}
	= D_{\mathrm{oxy}}\nabla^2 C_{\mathrm{oxy}}
	- \gamma\,c\,C_{\mathrm{oxy}},
	\label{C_Oxygen}
\end{equation}
	where \(D_{\mathrm{oxy}}=2000\,\mu\mathrm{m}^2\,\mathrm{s}^{-1}\)~\cite{ferrel1967diffusion} is the molecular diffusivity of oxygen in water and \(\gamma=10^{-5}\,\mathrm{mm}^3\,\mathrm{s}^{-1}\) is the assumed per-cell oxygen consumption rate.  We then modulate the swimming speed \(W_c\) based on the local oxygen concentration, increasing \(W_c\) with oxygen levels via a sigmoidal function~\cite{fragkopoulos2021self}:
\begin{equation}
	W_c(C_{\mathrm{oxy}})
	= W_{\min}
	+ \frac{W_{\max}-W_{\min}}{2}
	\Bigl[1 + \tanh\!\bigl(\tfrac{C_{\mathrm{oxy}} - C_{\mathrm{typ}}}{\beta}\bigr)\Bigr],
	\label{W_c_Oxygen}
\end{equation}
	with parameters
	\[
	W_{\min}=56\,\mu\mathrm{m\,s}^{-1},\quad
	W_{\max}=112\,\mu\mathrm{m\,s}^{-1},
	\]
	and
	\[
	C_{\mathrm{sat}}=3\times10^{-4}\,\mathrm{\mu mol\,mm}^{-3},\quad
	C_{\mathrm{typ}}=0.14\,C_{\mathrm{sat}},\quad
	\beta=0.11\,C_{\mathrm{sat}}.
	\]
	Here, \(C_{\mathrm{sat}}\)  denotes the saturation concentration of oxygen in water, and \(C_{\mathrm{typ}}\) is a representative (typical) oxygen concentration. Simulations assume an initially uniform saturation, \(C_{\mathrm{oxy}}(\mathbf{x},0)=C_{\mathrm{sat}}\). 
\begin{figure}[!t]
	\begin{center}
		\includegraphics[width=\columnwidth]{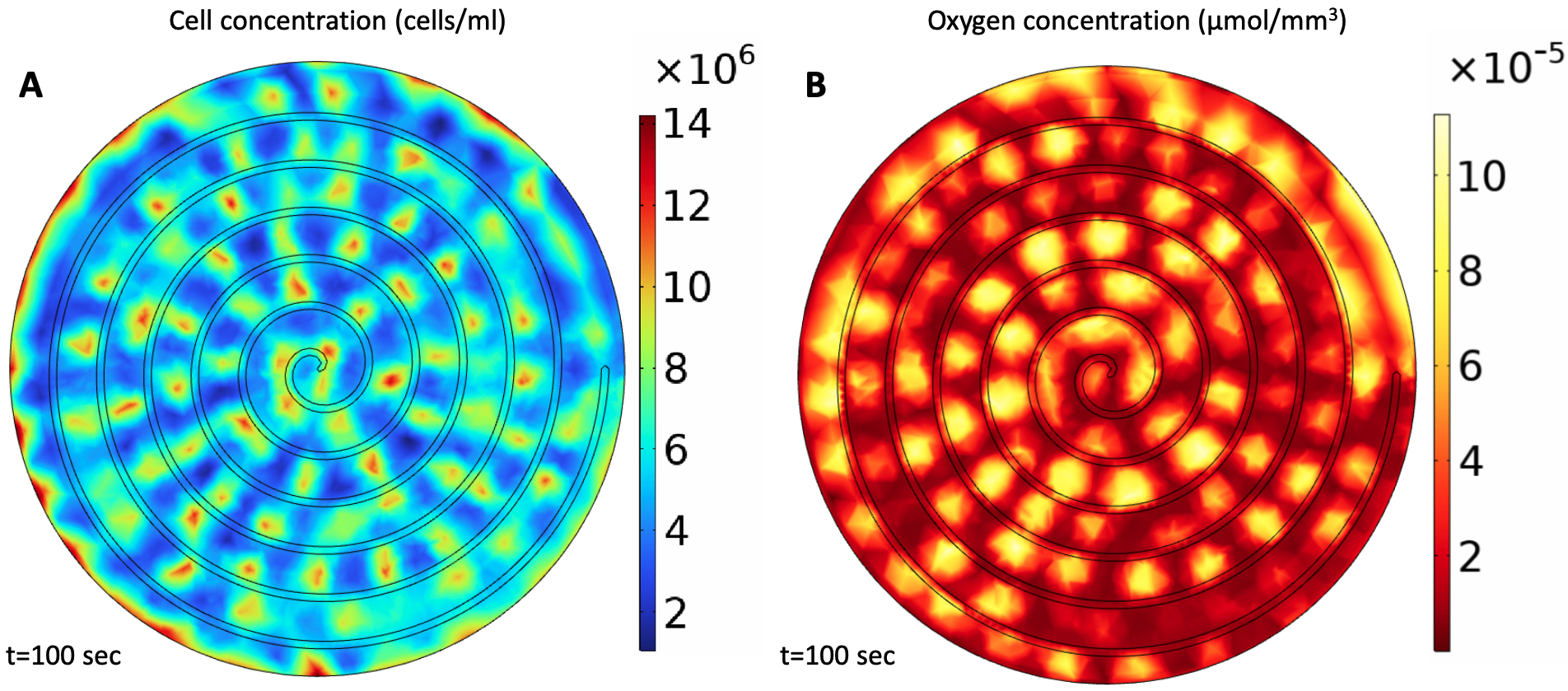}
		\caption{\textcolor{black}{Top-view snapshots at $z=3\ \mathrm{mm}$ from 3D simulations of Eqs.~\eqref{Eq:incompressibility}--\eqref{W_c_Oxygen} illustrate an anti-correlation between cell and oxygen concentrations; see Video 15 for the full dynamic evolution.
		}} 
		\label{fig:Oxygen}
	\end{center}
\end{figure}

Figure \ref{fig:Oxygen} and Video 15 show the results of our simulations of Eqs. \eqref{Eq:incompressibility}–\eqref{W_c_Oxygen} using the parameters listed above, while all other parameters remain as in our previous simulations. We imposed a no-flux boundary condition for oxygen concentration and a no-slip condition for velocity on all domain boundaries . As expected, cell and oxygen concentrations are anti-correlated: regions with higher cell density exhibit lower oxygen levels. As oxygen levels fall, swimming speed declines, ultimately reaching the minimum value of $56\,\mu\mathrm{m}\,\mathrm{s}^{-1}$ prescribed by Eq.~\eqref{W_c_Oxygen}. However, our simulations indicate that oxygen dynamics alone cannot reproduce the dramatic phase transitions observed in our experiments, underscoring the necessity of more sophisticated models to accurately capture these complex transitions. Moreover, potential crowding effects at elevated cell densities should also be considered to fully account for the intricate interplay between metabolic activity, confinement, and collective flow phenomena observed in our system.}

In summary, our study highlights the pivotal role of metabolic activity and oxygen dynamics in governing collective microbial behavior and demonstrates how environmental design—through boundary geometry and oxygen control—can be leveraged to manipulate microbial self-organization and fluid transport in active matter systems. More broadly, these findings offer new strategies for directing microbial collective flows and structuring living fluids through targeted environmental and geometric interventions, with potential applications in bioreactors, synthetic ecosystems, and the design of bioinspired active materials.\\
%
\\
\noindent\textbf{Acknowledgements} \par 
\noindent We sincerely thank Professors L. Dede, G. Shubeita, A. Narayanan, M. Ali, A. Pumir, H. de Maleprade, F. Picella, E. Bodenschatz and S. Ramaswamy for their invaluable insights and stimulating discussions. We are also grateful to Dr. Malavath for his dedicated support. Special thanks are extended to the Core Technology Platform (CTP) at NYU Abu Dhabi for their exceptional assistance, particularly to V. Dhanvi and J. Govindan for their outstanding work in fabricating the acrylic molds. Finally, we acknowledge the High Performance Computing (HPC) facility at NYU Abu Dhabi for providing critical computational resources and data storage support essential to this study.
 \\

\noindent\textbf{Author contributions} \par
\noindent S.G., S.O.A. and T.D. conducted the experiments, I.G., S.V.R.A., A.B. and A.G. designed and carried out the numerical simulations. A.G. conceived the idea, designed the experiments, analyzed the data, and wrote the manuscript.\\

\noindent\textbf{Data and materials availability} 
\noindent All data needed to evaluate the conclusions in the paper are present in the paper and/or the Supplementary Materials. Additional data related to this paper may be requested from the authors.
\bibliography{Bib} 
\bibliographystyle{rsc} 
\newpage
\section{Supplemental Figures}
\setcounter{equation}{0} 
\renewcommand{\theequation}{S.\arabic{equation}}
\renewcommand\thefigure{S\arabic{figure}}
\renewcommand\thepage{S\arabic{page}}
\setcounter{figure}{0}
\setcounter{page}{1}

\begin{figure}[!htbp]
	\begin{center}
		\includegraphics[width=\columnwidth]{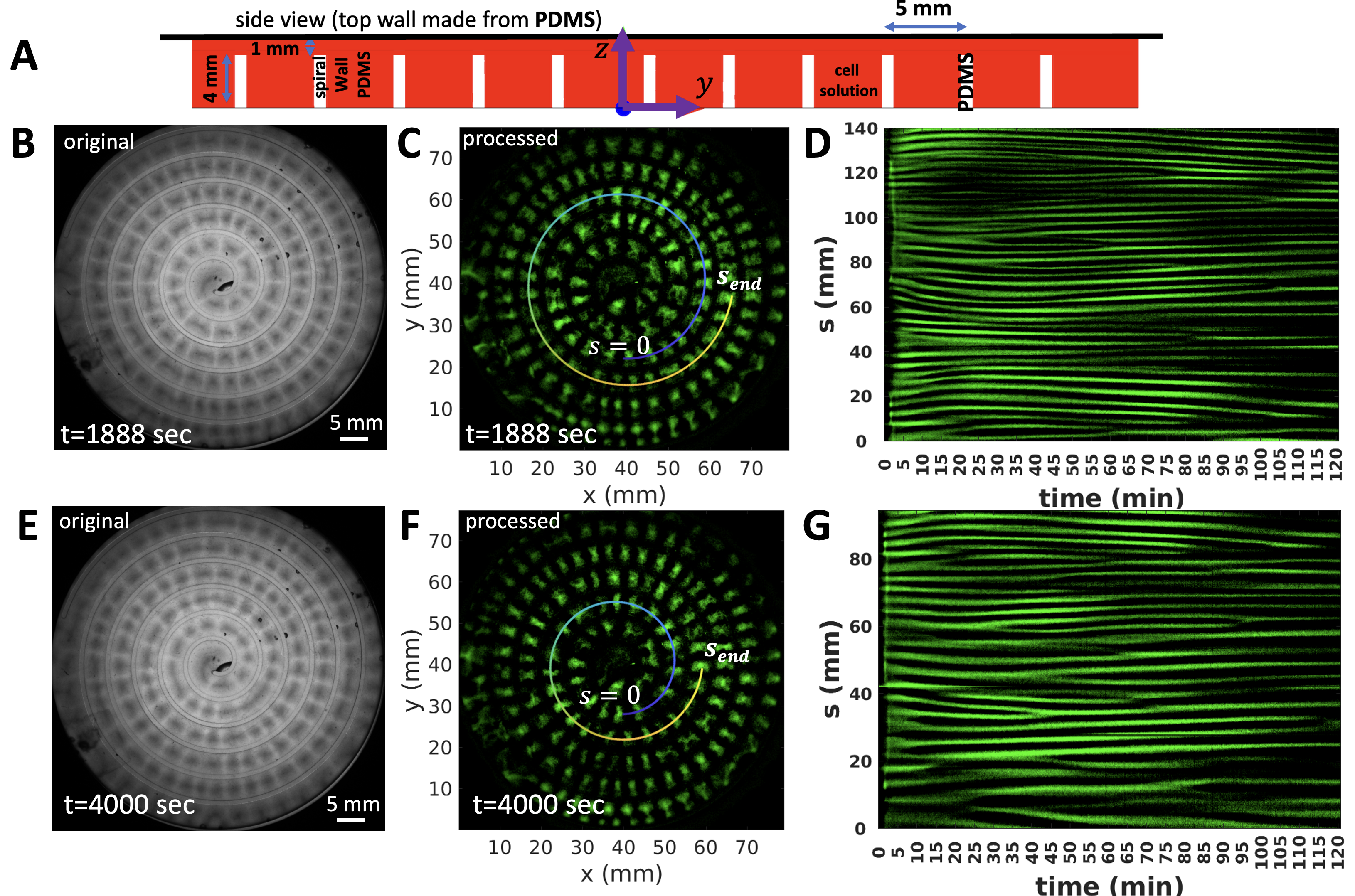}
		\caption{ A control experiment was performed using a fully confined setup fabricated entirely from air-permeable PDMS. As illustrated in the corresponding space–time plots, no transition in the bioconvection patterns was observed even after two hours—markedly contrasting the transitions seen in confined PMMA-based configurations. This result underscores the critical role of oxygen availability in driving the observed pattern transitions; see Video 16.} 
		\label{fig:All_PDMS}
	\end{center}
\end{figure}
\begin{figure}[!htbp]
	\begin{center}
		\includegraphics[width=\columnwidth]{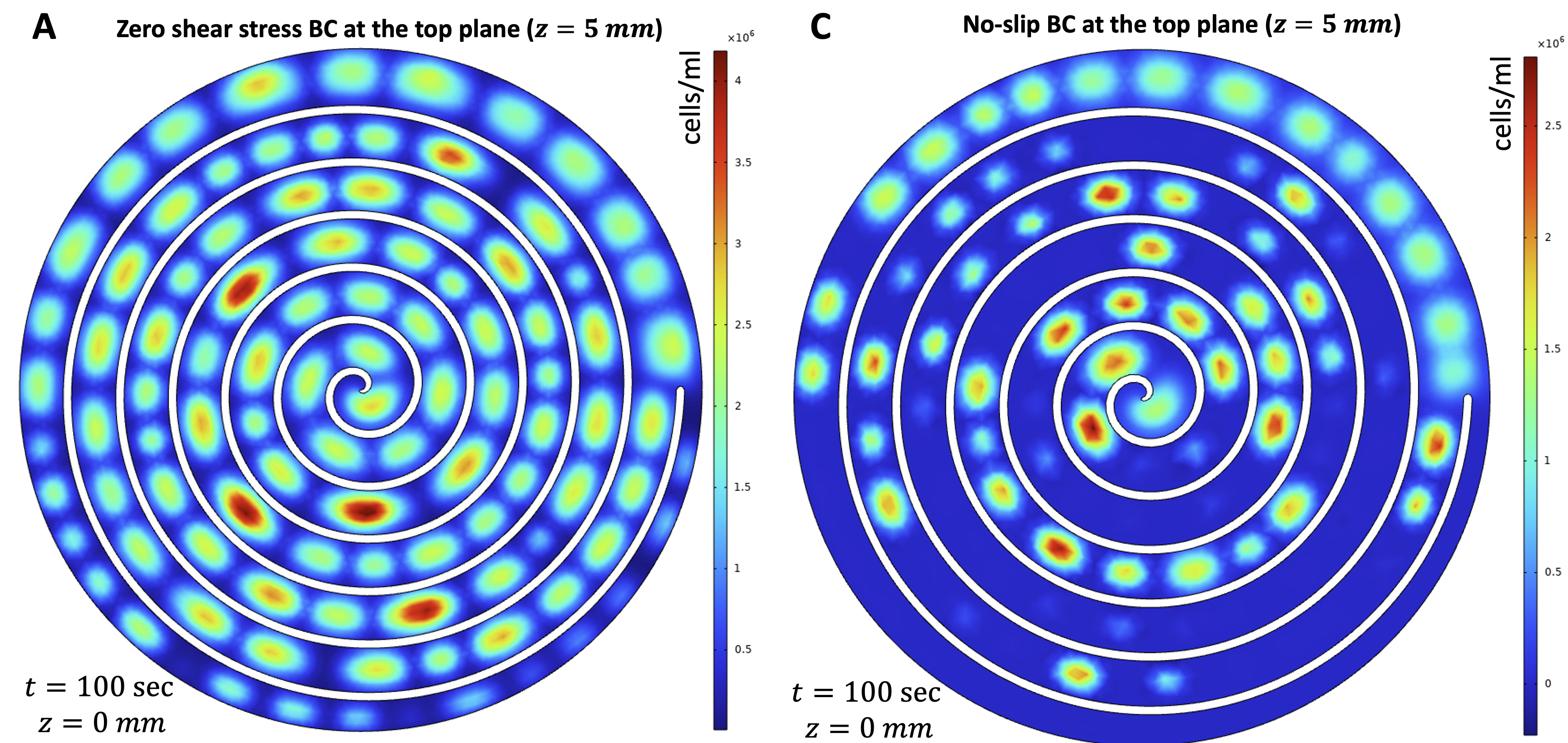}
		\includegraphics[width=\columnwidth]{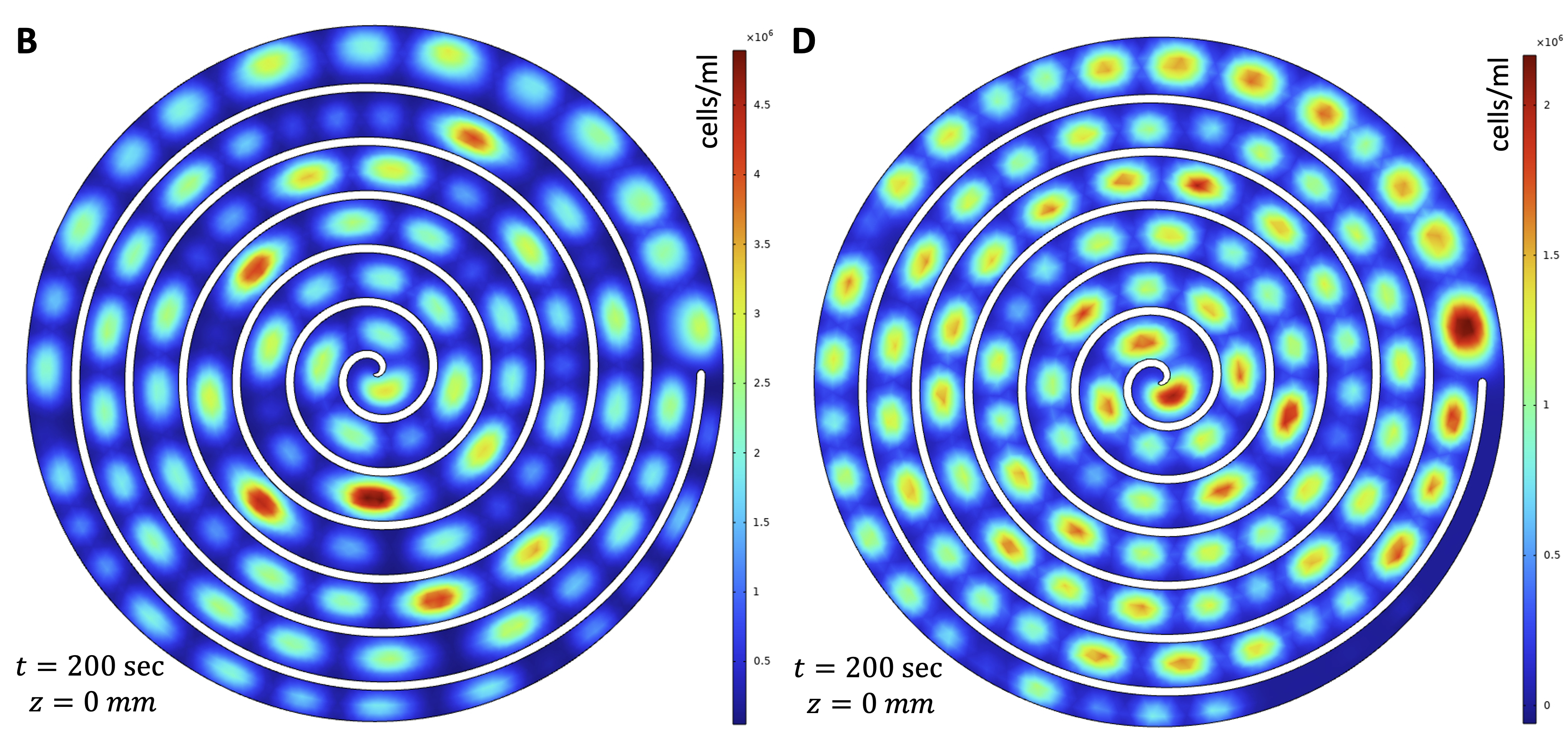}
		\caption{Cell density distribution at the bottom of the setup ($z = 0$) from 3D simulations of bioconvection patterns. Snapshots at $t = 100$ s reveal that the cell concentration at the bottom is approximately half that near the surface (see Fig.~\ref{fig:Simulations_Z4mm}), and the cell clusters remain more diffuse and unlike the near-surface structures, clusters at the bottom do not exhibit reorientation perpendicular to the spiral boundaries, highlighting the depth-dependent development of bioconvection patterns; see Videos 9 and 10 .} 
		\label{fig:Simulations_Z0mm}
	\end{center}
\end{figure}
\begin{figure}[t]
	\begin{center}
		\includegraphics[width=\columnwidth]{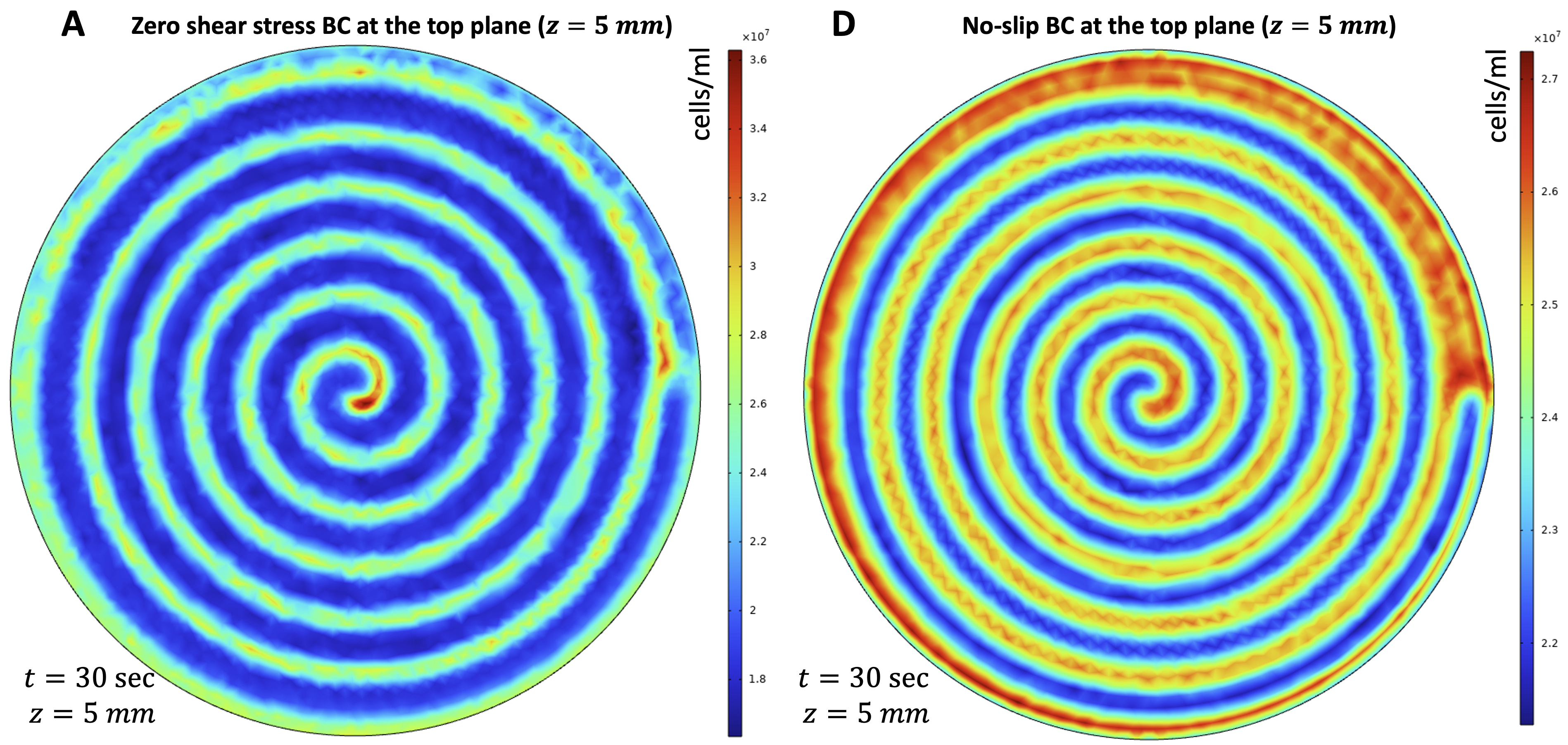}
		\includegraphics[width=\columnwidth]{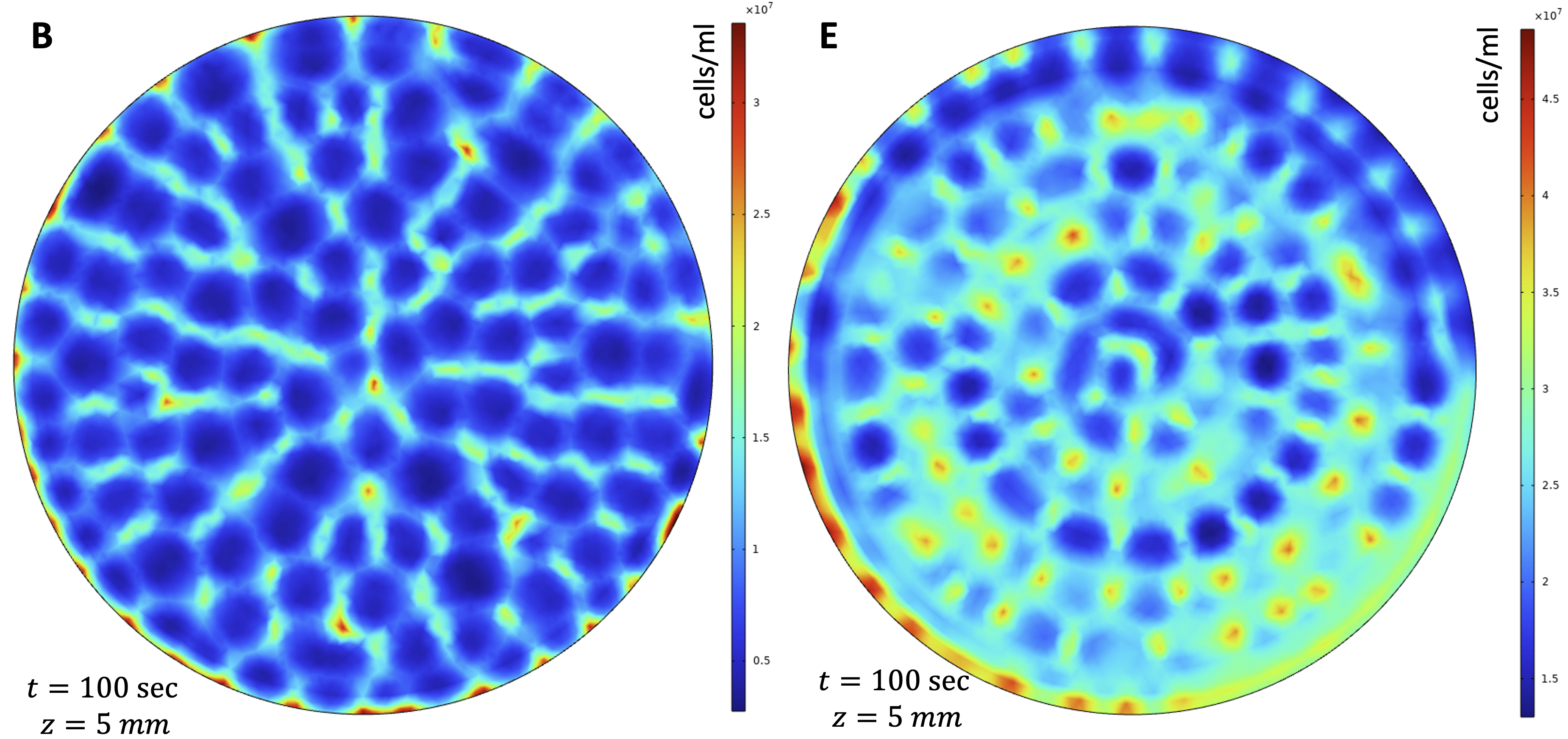}
		\includegraphics[width=\columnwidth]{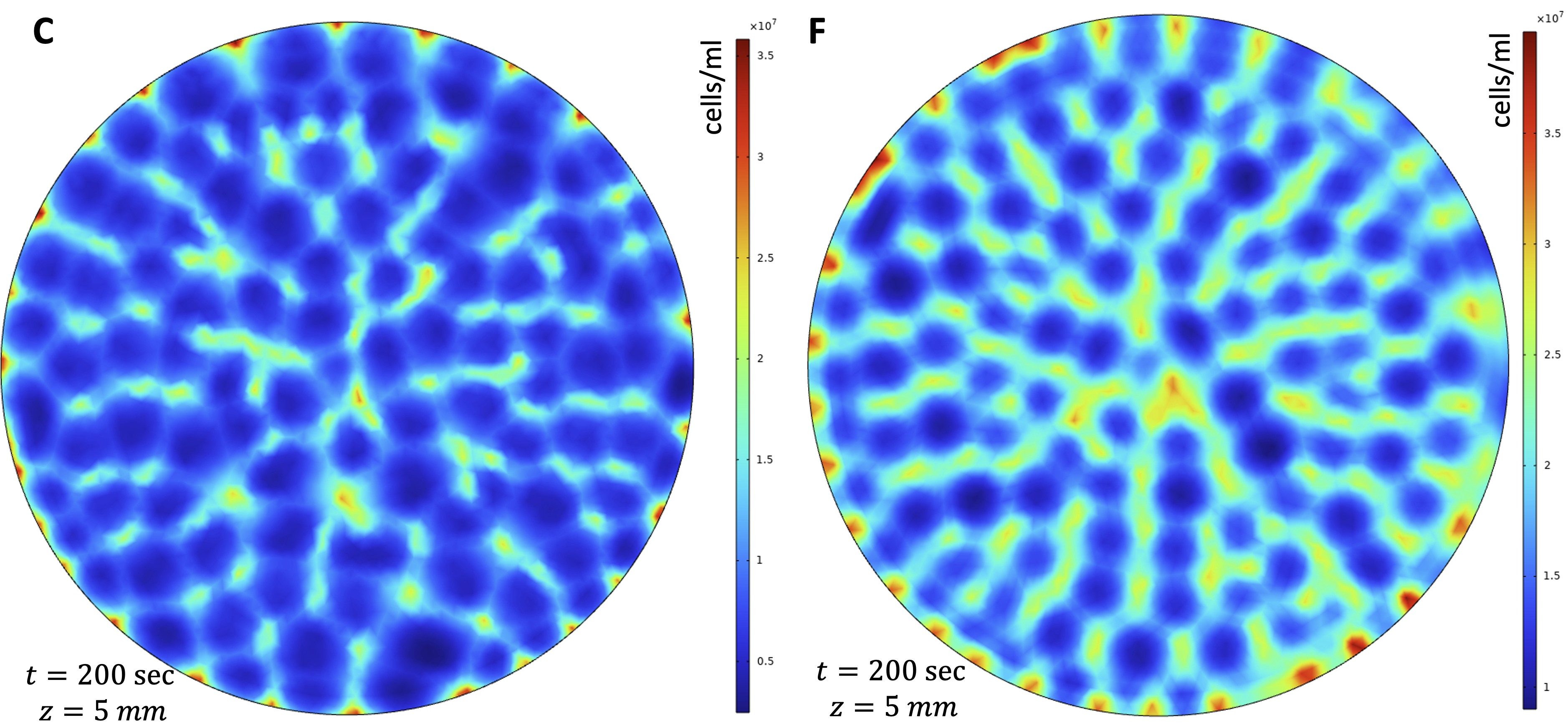}
		\caption{ Top-view snapshots of cell density from the same 3D simulations presented in Fig.~\ref{fig:Simulations_Z4mm}, but taken 1\, mm above the spiral-shaped boundary at $z = 5$ mm. This plane corresponds to the fluid-air interface in the left panels and to the rigid top wall boundary in the right panels. Either a zero-shear-stress condition (representing a free liquid-air interface) or a no-slip condition (representing a rigid top wall) is applied. As described in Fig.~\ref{fig:Simulations_Z4mm}, \textit{CR} cells swim upward (negative gravitaxis) and accumulate near the surface, forming a distinct spiral-shaped cell density pattern (A, D). This structure is dynamically unstable, fragmenting into smaller, high-density clusters that subsequently evolve into downwelling plumes; see also Fig.~\ref{fig:Flow} and Video 17.} 
		\label{fig:Simulations_Z5mm}
	\end{center}
\end{figure}
\end{document}